\documentclass[%
 reprint,
 amsmath,amssymb,
 aps,
prb,
]{revtex4-1}
	\usepackage{feynmf}
	\usepackage{graphics}
	\usepackage{epsfig}
	\usepackage{graphicx}
	\usepackage{color}
	\usepackage{comment}
\begin{document}
	\unitlength = 1mm
	
	\title{Magnetic Couplings in Edge-Sharing High-Spin $d^7$ Compounds}
\date{\today}
\begin{abstract}  
High-spin $d^7$ Co(II) compounds have recently been identified as possible platforms for realising highly anisotropic and bond-dependent couplings featured in quantum-compass models such as the celebrated Kitaev model. In order to evaluate this potential, we consider all symmetry-allowed contributions to the magnetic exchange for ideal edge-sharing bonds. Though a combination of ab-initio and cluster many-body calculations we conclude that bond-dependent couplings are generally suppressed in favor of Heisenberg exchange for real materials. Consequences for several prominent materials including Na$_2$Co$_2$TeO$_6$ and BaCo$_2$(AsO$_4$)$_2$ are discussed. 
\end{abstract}

\author{Stephen M. Winter*}
\affiliation{Department of Physics and Center for Functional Materials, Wake Forest University, NC 27109, USA}

\maketitle

\section{Introduction}

Pursuit of strongly anisotropic $d$-block magnets has been motivated by the possibility of material realization of quantum compass models\cite{nussinov2015compass}, such as Kitaev's celebrated honeycomb model\cite{kitaev2006anyons}. 
In these materials, competition between different bond-dependent magnetic interactions produces an extensive classical degeneracy conducive to quantum spin-liquid ground states\cite{hermanns2018physics,broholm2020quantum,zhou2017quantum}. Realising these conditions in real materials requires precise tuning and suppression of the usual isotropic magnetic exchange. This can be accomplished, in principle, in edge-sharing compounds with $d^5$ filling and strong spin-orbital coupling. Remarkably, for ideal considerations, the specific spin-orbital composition of the local moments suppresses all couplings except those bond-dependent Ising couplings precisely prescribed by the Kitaev model\cite{jackeli2009mott}. This revelation initiated a number of studies\cite{winter2017models,trebst2017kitaev} in $5d^5$ Ir(IV) compounds such as A$_2$IrO$_3$ (A = Na, Li)\cite{singh2010antiferromagnetic} and the $4d^5$ Ru(III) compound\cite{plumb2014alpha,banerjee2016proximate,banerjee2017neutron} $\alpha$-RuCl$_3$. These studies have revealed clear evidence of dominant bond-dependent anisotropic couplings in these compounds\cite{hwan2015direct,suzuki2021proximate}, leading to a variety of anomalous behaviors from the breakdown of conventional magnon excitations\cite{winter2017breakdown} to the possibility of a field-induced spin liquid\cite{banerjee2018excitations,kasahara2018majorana,yokoi2021half}. However, while Kitaev couplings are thought to be the largest interaction, other couplings of similar magnitude always lift the classical degeneracy leading to magnetic order at zero field.

In this context, the seminal work of Liu et al.\cite{liu2018pseudospin,doi:10.1142/S0217979221300061,liu2020kitaev} and Sano et al. \cite{sano2018kitaev} renewed hope for realizing Kitaev's spin liquid, by showing that the magnetic exchange in high-spin $3d^7$ Co(II) ions may also produce dominant Kitaev interactions for ideal considerations. In particular, these studies assumed the dominant hopping between transition metal ions occurs via hybridization with ligand ions, which suppresses other couplings. While this condition is satisfied for $5d^5$ Ir(IV) compounds such as A$_2$IrO$_3$, the presence of significant direct metal-metal hopping in $4d^5$ $\alpha$-RuCl$_3$ is the primary source of non-Kitaev interactions\cite{rau2014generic,winter2016challenges}. It is not clear that these assumptions are satisfied in $3d$ systems. 

The possibility of strong bond-dependent Kitaev interactions also challenges the conventional view that Co(II) compounds  typically have bond-independent XXZ anisotropy largely driven by the effects of local crystal field distortions on the $j_{1/2}$ doublets \cite{lines1963magnetic,oguchi1965theory}. For example, CoNb$_2$O$_6$ (CNO) is considered to be a prototypical 1D Ising ferromagnet\cite{scharf1979magnetic,maartense1977field,kobayashi1999three}, and has been studied in the context of transverse-field Ising criticality \cite{lee2010interplay,coldea2010quantum,morris2014hierarchy}. The structure consists of zigzag chains of edge-sharing CoO$_6$ octahedra. While this bonding geometry might be expected to produce large bond-dependent couplings, the dominant nearest neighbor interaction is known to have an Ising form $S_i^\alpha S_j^\alpha$ with a common $\alpha$-axis for all bonds.  Recent studies have highlighted the importance of small deviations\cite{fava2020glide,morris2021duality}, but it is nonetheless evident that the Kitaev coupling is not dominant. 

More recently, the pursuit of 2D honeycomb materials with large bond-dependent couplings has drawn attention to Na$_3$Co$_2$SbO$_6$ (NCSO), and Na$_2$Co$_2$TeO$_6$ (NCTO). Both materials show zigzag antiferromagnetic order\cite{PhysRevB.94.214416,PhysRevB.95.094424,wong2016zig,PhysRevB.95.094424,chen2021spin}. This ground state is natural for strong bond-dependent couplings\cite{chaloupka2013zigzag,rau2014generic}, although longer-range Heisenberg $J_2$ and $J_3$ may also be invoked\cite{fouet2001investigation,kimchi2011kitaev} Indeed, analysis of inelastic neutron scattering has led to a wide variety of proposed models for the couplings\cite{chen2021spin,songvilay2020kitaev,kim2021antiferromagnetic,lin2021field}, which span the entire range from dominant Heisenberg to dominant Kitaev. Overall, the relative role of nearest neighbor bond-dependent coupling vs.~longer range Heisenberg exchange remains unclear.

Two more honeycomb materials of recent interest are BaCo$_2$(AsO$_4$)$_2$ (BCAO) and BaCo$_2$(PO$_4$)$_2$ (BCPO). Of these, BCPO displays only short-range incommensurate correlations, suggesting strong frustration\cite{nair2018short}. BCAO orders in a state intermediary between zigzag antiferromagnetic and ferromagnetic states with unconventional magnon dispersion\cite{regnault2006investigation,regnault2018polarized}, which has been discussed as an incommensurate helimagnet\cite{regnault1977magnetic} or double stripe zigzag\cite{regnault2018polarized}. Under applied field in-plane, BCAO undergoes a series of phase transitions\cite{regnault1977magnetic} between magnetization plateaus, and was proposed to host a field-induced spin liquid\cite{zhong2020weak,zhang2021and}. However, this was recently called into question due to the appearance of sharp magnon modes in each of the phases\cite{shi2021magnetic}. As with NCTO, the relative role of different couplings is a subject of much discussion; the first {\it ab-initio} studies\cite{das2021xy,maksimov2022} favored a nearly XXZ model, in contrast with the assumption of large Kiteav interactions.

All of these findings call for a reinvestigation of the magnetic couplings in edge-sharing Co(II) materials. In this work, we find, in contrast to the assumptions of the initial theoretical analysis, that ligand-mediated hopping is {\it not} large in these compounds. For this reason the character of the magnetic couplings is significantly altered from the expected Kitaev form. In particular, ferromagnetic Heisenberg $J$ typically dominates the nearest neighbor couplings, while a variety of smaller anisotropic couplings may appear depending on the specific details of the hopping and crystal field distortions, which are discussed in this work.

The paper is organized as follows: We first define the electronic Hamiltonian, and review the single-ion ground state, and the effect of crystal field distortions on the spin-orbital composition of the $j_{1/2}$ moments. We then analyze the full set of relevant symmetry-allowed hoppings in edge-sharing bonds, and present calculations of such hoppings for real materials to establish their physically relevant ranges. On the basis of these hoppings, we then compute the resulting magnetic couplings. Finally, we summarize the results, and discuss them in the context of materials of recent interest.

\section{Electronic Hamiltonian}

\subsection{General Form}
\begin{figure}[t]
\includegraphics[width=\linewidth]{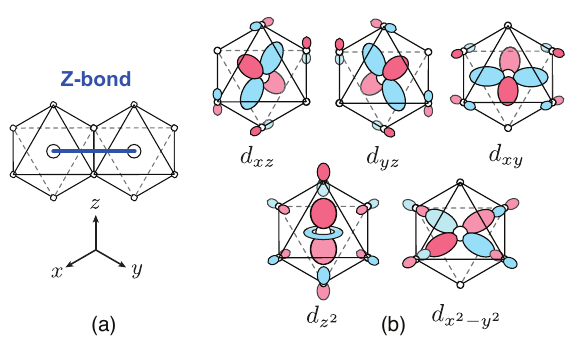}
\caption{(a) Edge-sharing bond with cubic coordinates defined. (b) Cartoon of the $d$-orbital Wannier functions in the {\it downfolded} $d$-only basis.}
\label{fig:wannier}
\end{figure}

We first define the electronic Hamiltonian relevant for the edge-sharing bond depicted in Fig.~\ref{fig:wannier}(a), which is conventionally called a Z-bond to indicate that the global cubic $z$-axis is perpendicular to the shared edge. There are generally two different orbital bases that can be employed in the description of magnetic couplings. One may consider explicitly either (i) the full basis of metal \mbox{$d$-orbitals} and ligand $p$-orbitals, or (ii) the \mbox{$d$-orbitals} only, for which the effects of the $p$-orbitals are {\it downfolded} into the \mbox{$d$-orbital} Wannier function basis. The latter Wannier functions are anti-bonding combinations of $d$- and $p$-orbitals depicted schematically in Fig.~\ref{fig:wannier}(b). In the following, we mostly consider the {\it downfolded} basis, but discuss the deficiencies of this approach below. 

In the downfolded $d$-only basis, the total Hamiltonian is given by:
\begin{align}\label{eqn:overallham}
\mathcal{H} = \mathcal{H}_{\rm hop} + \sum_i \mathcal{H}_i
\end{align}
where $\mathcal{H}_{\rm hop}$ denotes the hopping between Co sites. $\mathcal{H}_i$ is the local Hamiltonian at each Co site $i$, which is a sum, respectively, of Coulomb interactions, crystal-field splitting, and spin-orbit coupling:
\begin{align}
\mathcal{H}_i = \mathcal{H}_{U} + \mathcal{H}_{\rm CFS} + \mathcal{H}_{\rm SOC}
\label{eqn_hi}
\end{align}
The SOC term is written:
\begin{align}
\mathcal{H}_{\rm SOC} = \lambda \mathbf{L}_i \cdot \mathbf{S}_i
\end{align}
where the atomic SOC constant for Co is roughly $\lambda_{\rm Co} \approx 60$ meV.  
The Coulomb interactions are most generally written:
\begin{align}
\mathcal{H}_U = \sum_{\alpha,\beta,\delta,\gamma}\sum_{\sigma,\sigma^\prime}U_{\alpha\beta\gamma\delta} \ c_{i,\alpha,\sigma}^\dagger c_{i,\beta,\sigma^\prime}^\dagger c_{i,\gamma,\sigma^\prime} c_{i,\delta,\sigma}
\end{align}
where $\alpha,\beta,\gamma,\delta$ label different $d$-orbitals. 
\footnote{The coefficients $U_{\alpha\beta\gamma\delta}$ may be grouped according to the number of unique orbital indices, from one to four. For example, the intra-orbital Hubbard terms $n_{i,\alpha,\uparrow}n_{i,\alpha,\downarrow}$ have one unique index $\alpha$, while the inter-orbital Hubbard terms $n_{i,\alpha,\sigma}n_{i,\beta,\sigma^\prime}$ have two unique indices $\alpha,\beta$. In the spherically symmetric approximation \cite{sugano2012multiplets}, the Coulomb coefficients with three and four indices vanish unless at least one of the orbitals is an $e_g$ orbital. For this reason, $t_{2g}$-only (and $e_g$-only) models reduce to the familiar Kanamori form\cite{georges2013strong,pavarini2014dmft}, which includes only Hubbard density-density repulsion, Hund's exchange, and pair-hopping contributions. However, when both $e_g$ and $t_{2g}$ orbitals are considered together, it is important to include the full rotationally symmetric Coulomb terms. This is particularly true when computing anisotropic magnetic exchange, because any approximations to the Coulomb Hamiltonian are likely to explicitly break rotational symmetry, leading to erroneous sources of anisotropy.}
In the spherically symmetric approximation \cite{sugano2012multiplets}, the coefficients $U_{\alpha\beta\gamma\delta}$ are all related to the three Slater parameters $F_0, F_2, F_4$. In terms of these, the familiar $t_{2g}$ Kanamori parameters are, for example:
\begin{align}
U_{t2g} = F_0 + \frac{4}{49} \left( F_2 + F_4 \right) \\
J_{t2g} = \frac{3}{49} F_2 + \frac{20}{441} F_4
\end{align}
Unless otherwise stated, we use $U_{t2g} = 3.25$ eV, and $J_{t2g} = 0.7$ eV to model Co(II) compounds, following Ref.~\onlinecite{das2021xy}. We also take the approximate ratio $F_4/F_2 = 5/8$, which is appropriate for $3d$ elements\cite{pavarini2014dmft}. 

\begin{figure}[t]
\includegraphics[width=0.75\linewidth]{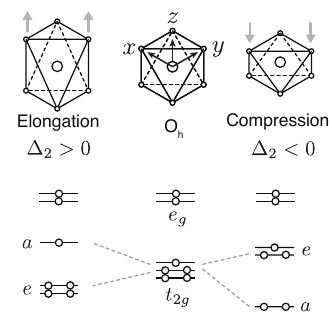}
\caption{Energy level diagram showing the splitting of the local single-particle electronic orbitals in the absence of spin-orbit coupling. Occupied orbitals in the high-spin $d^7$ configuration are indicated.}
\label{fig:coords}
\end{figure}

For the crystal-field Hamiltonian, we consider an ideal trigonal distortion within $D_{3d}$ site symmetry. The Hamiltonian can be written:
\begin{align}
\mathcal{H}_{\rm CFS} = \sum_{\sigma} \mathbf{c}_{i,\sigma}^\dagger \ \mathbb{D} \ \mathbf{c}_{i,\sigma}
\end{align}
where:
\begin{align}
\mathbf{c}_{i,\sigma}^\dagger = \left(c_{i,yz,\sigma}^\dagger \  c_{i,xz,\sigma}^\dagger \ c_{i,xy,\sigma}^\dagger \ c_{i,z^2,\sigma}^\dagger \ c_{i,x^2-y^2,\sigma}^\dagger\right)
\end{align}
creates an electron in one of the $d$-orbital (Wannier) functions at site $i$. 
In terms of the global cubic $(xyz)$ coordinates defined in Fig.~\ref{fig:coords}, the CFS matrix can be written:
\begin{align}
\mathbb{D} = \left( \begin{array}{ccccc} 
0 & \Delta_2 & \Delta_2 & 0 & 0 \\
\Delta_2 & 0 & \Delta_2 & 0 & 0 \\
\Delta_2 & \Delta_2 & 0 & 0 & 0 \\
0 & 0 & 0 & \Delta_1 & 0 \\
0 & 0 & 0 & 0 & \Delta_1
\end{array}\right)
\end{align}
where $\Delta_1$ is the $t_{2g}$-$e_g$ splitting, and $\Delta_2$ is the trigonal term. Trigonal distortions are relevant for BCAO, BCPO, NCTO, and NCSO. Generally, $\Delta_2>0$ corresponds to trigonal elongation, as shown in Fig.~\ref{fig:coords}, although the actual sign is further influenced by the details of the ligand environments and longer ranged Coulomb potentials. Without SOC, the $t_{2g}$ levels are split into a doubly degenerate $e$ pair and a singly degenerate $a$ level, with $E_{a} - E_{e} = 3\Delta_2$. 
As discussed below, the trigonal splitting has a strong impact on the nature of the local moments. 

\begin{figure}[t]
\includegraphics[width=0.9\linewidth]{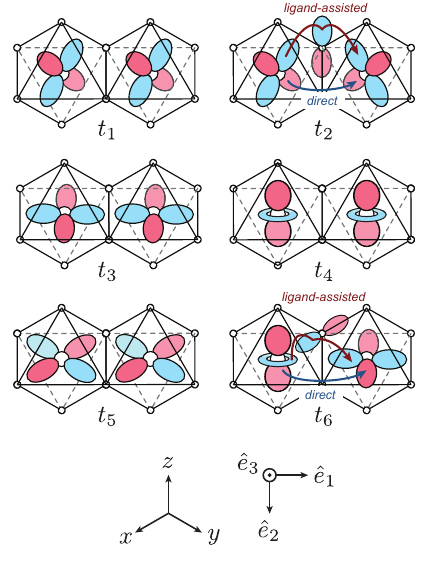}
\caption{Summary of symmetry allowed hoppings for ideal Z-bonds with $C_{2v}$ symmetry and $90^\circ$ metal-ligand-metal bond angles. $t_1, t_3, t_4$, and $t_5$ arise from direct metal-metal hopping, while $t_2$ and $t_6$ have contributions from both direct and ligand-assisted processes. The global $(xyz)$ and local $(\hat{e}_1 \hat{e}_2 \hat{e}_3)$ coordinates are shown. The $p$-orbital contributions to the Wannier functions have been omitted except in cases where ligand-assisted hopping is depicted.}
\label{fig:hops}
\end{figure}

The effective $d$-$d$ hopping between transition metal sites is described by:
\begin{align}
\mathcal{H}_{\rm hop} = \sum_{ij,\sigma} \mathbf{c}_{i,\sigma}^\dagger \ \mathbb{T}_{ij} \ \mathbf{c}_{j,\sigma}
\end{align}
For an ideal edge-sharing bond with $90^\circ$ metal-ligand-metal bond angles, $C_{2v}$ symmetry restricts the form of the hopping matrices. In terms of the global $(x,y,z)$ coordinates defined in Fig.~\ref{fig:hops}, the matrices are constrained to take the following form, for the Z-bond:
\begin{align}
\mathbb{T}_{Z} = \left( \begin{array}{ccccc} 
t_1 & t_2 & 0 & 0 & 0 \\
t_2 & t_1 & 0 & 0 & 0 \\
0 & 0 & t_3 & t_6 & 0 \\
0 & 0 & t_6 & t_4 & 0 \\
0 & 0 & 0 & 0 & t_5
\end{array}\right)
\end{align}
Of these, $t_1, t_3, t_4$, and $t_5$ are primarily direct hopping between the $d$-orbitals on the metal atoms, as shown in Fig.~\ref{fig:hops}. By contrast,  $t_2$ and $t_6$ have significant contributions from both direct hopping and ligand-assisted hopping, which arises from the hybridization between the metal and ligand orbitals. Realistic ranges for these hoppings are discussed in section \ref{sec:hops}.

Previous works\cite{liu2018pseudospin,doi:10.1142/S0217979221300061,liu2020kitaev,sano2018kitaev} on the possibility of dominant Kitaev magnetic exchange in high-spin $3d^7$ Co(II) ions have highlighted the ideal regime is found for small trigonal splitting ($\Delta_2\ll \lambda$) and dominant ligand-assisted hopping ($t_2 \gg t_1, t_3, t_4, t_5, t_6$). In principle, if either of these conditions is violated, the interactions may deviate significantly from the Kitaev form. We consider these possible deviations in the following sections.

\subsection{Composition of Local Moments}

For the ``high-spin'' $d^7$ case, the ground state has nominal configuration $(t_{2g})^5 (e_g)^2$, with three unpaired electrons ($S = 3/2$), as shown in Fig.~\ref{fig:coords}. For $3d$ transition metals, the constants in the Hamiltonian $\mathcal{H}_i$ typically satisfy $J_H, \Delta_1 \gg \lambda,\Delta_2$. This ensures that configurations belonging {\it precisely} to the $(t_{2g})^5 (e_g)^2$, $S = 3/2$ manifold carry the dominant weight in the low energy multiplets, even when SOC and trigonal splitting are considered. For this reason, the low-energy single-ion states may be conveniently described by {\it projecting} $\mathcal{H}_i$ into this manifold, and diagonalizing. This procedure was employed in Ref.~\onlinecite{lines1963magnetic}, and is repeated in this section for the purpose of discussion. More detailed discussions can be found in Ref.~\onlinecite{lines1963magnetic, doi:10.1142/S0217979221300061}. 

In the absence of trigonal splitting ($\Delta_2 = 0$), there is a three-fold orbital degeneracy associated with the $t_{2g}$ levels, leading to an effective orbital momentum $L_{\rm eff} = 1$. Spin-orbit coupling $\mathcal{H}_{\rm SOC} = \lambda \mathbf{L}\cdot \mathbf{S}$ splits the low energy multiplets into $J_{\rm eff} = 1/2$, 3/2, and $5/2$ states. For $\Delta_2 = 0$, the multiplet energies satisfy:
\begin{align}
E_{3/2} - E_{1/2} = \frac{1}{2} \lambda
\\
E_{5/2} - E_{1/2} = \frac{4}{3} \lambda
\end{align}
Given $\lambda_{\rm Co}  \approx 60$ meV, the $j_{1/2} \to j_{3/2}$ excitation is expected to appear in the range of $\sim30$ meV, as has been seen experimentally in numerous compounds\cite{PhysRevB.98.024415,ross2017single,songvilay2020kitaev,kim2021antiferromagnetic}. 

\begin{figure}[t]
\includegraphics[width=0.8\linewidth]{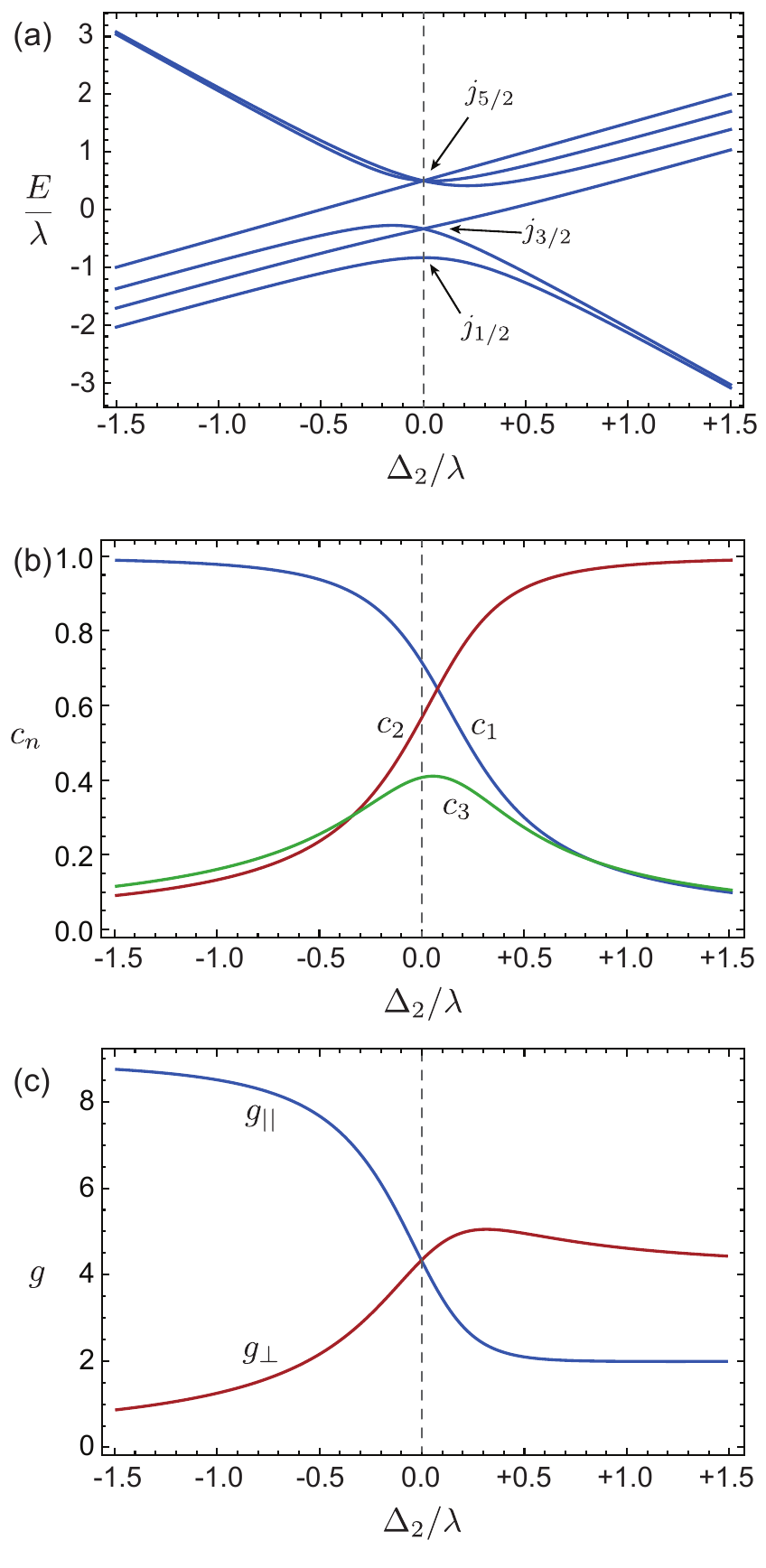}
\caption{Evolution of the single-ion properties with trigonal CFS $\Delta_2$ following Ref.~\onlinecite{lines1963magnetic}. (a) Energy spectrum. (b) Wavefunction coefficients $c_n$ for the ground state $j_{1/2}$ doublet. (c) $g$-tensor components for the ground state doublet. $g_{||}$ refers to the direction parallel to the trigonal distortion axis, $\hat{x}+\hat{y}+\hat{z}$.}
\label{fig:single}
\end{figure}

For finite $\Delta_2$, the evolution of the multiplet energies is shown in Fig.~\ref{fig:single}(a). For all values of trigonal splitting, the single-ion ground state is a doublet; it is smoothly connected to the pure $J_{\rm eff} = 1/2$ doublet as $\Delta_2$ vanishes. 
The ground state doublet can be written in terms of $|m_L, m_S\rangle$ states as\cite{lines1963magnetic}:
\begin{align}
\left|j_{1/2},+\frac{1}{2}\right\rangle = & \ c_1 \left|-1,\frac{3}{2}\right\rangle + c_2 \left|0,\frac{1}{2}\right\rangle + c_3 \left|1,-\frac{1}{2}\right\rangle \label{eqn:j12a} \\
\left|j_{1/2},-\frac{1}{2}\right\rangle = & \ c_1 \left|1,-\frac{3}{2}\right\rangle + c_2 \left|0,-\frac{1}{2}\right\rangle + c_3 \left|-1,\frac{1}{2}\right\rangle \label{eqn:j12b}
\end{align}
where the coefficients $c_n$ vary with $\Delta_2, \lambda$ as shown in Fig.~\ref{fig:single}(b). More detailed expressions for the $|m_L, m_S\rangle$ states in terms of the occupation of the single-particle $d$-orbitals is given in Appendix \ref{sec:appendixA}. Analytical expression for $c_n$ can be found in Ref.~\onlinecite{lines1963magnetic}; for $\Delta_2 = 0$, the coefficients are $c_1 = 1/\sqrt{2}$, $c_2 = 1/\sqrt{3}$, $c_3 = 1/\sqrt{6}$. From Fig.~\ref{fig:single}(b), it is evident that the composition of the lowest doublet changes dramatically with finite $\Delta_2$.

In the limit of large trigonal elongation $\Delta_2 > 0 $, the wavefunction coefficients converge to $c_2 \to 1$ and $c_1,c_3 \to 0$. This reflects the fact that the unpaired hole in the $t_{2g}$ levels would occupy the singly degenerate $a$ level, thus quenching the orbital moment completely (see Fig.~\ref{fig:coords}). As a result, spin-orbit coupling effects are suppressed, and the fourfold degeneracy of the nearly pure $S = 3/2$ moment is restored, such that the energetic splitting between the lowest two doublets becomes small (Fig.~\ref{fig:single}(a)). In this limit, the $m_s = \pm 1/2$ states lie slightly below the $m_s = \pm 3/2$ states due to the residual effects of SOC, and thus the splitting of the lowest two doublets may be considered as a residual easy-plane single-ion anisotropy. As such, the $g$-tensor (Fig.~\ref{fig:single}(c)) for the lowest doublet satisfies $g_\perp > g_{||}$, where $g_{||}$ refers to the component along the trigonal $(111)$ axis. 

For the opposite case of trigonal compression $\Delta_2 < 0 $, the unpaired hole in the $t_{2g}$ levels occupies the doubly degenerate $e$ levels (see Fig.~\ref{fig:coords}), thus retaining some orbital degeneracy consistent with $L_{\rm eff} = 1/2$. As depicted in Fig.~\ref{fig:single}(a), the effect of SOC is then to split the $S = 3/2$, $L_{\rm eff} = 1/2$ manifold into four doublets. The lowest doublet remains energetically separated from the higher states. For increasing trigonal distortion, the wavefunction coefficients for the lowest doublet approach $c_1 \to 1$, $c_2,c_3 \to 0$, implying that it is composed of pure $m_s = \pm 3/2$ states, according to Eq.~(10, 11). Consistently, the $g$-tensor satisfies $g_{||} \gg g_{\perp}$ in this limit (Fig.~\ref{fig:single}(c)).

Finally, it should be emphasized that an effective \mbox{$J = \frac{1}{2}$} model that incorporates only the lowest doublet remains valid only as long as the energetic splitting between the lowest doublet and first excited state remains large compared to the intersite magnetic exchange. For large $\Delta_2 < 0$, the gap between the lowest doublets converges to $\lambda/3 \sim 20$ meV, which should typically exceed the intersite magnetic coupling. For this reason, a model incorporating only the lowest doublet may remain valid even for large $\Delta_2 < 0$. For $\Delta_2 > 0$, which is expected to be relevant for BCAO, BCPO, NCTO, and NCSO, the range of validity is roughly constrained to $\Delta_2 < \lambda/2$. For larger values of $\Delta_2$, it would be more appropriate to consider an $S = 3/2$ model with easy-plane single-ion anisotropy. A detailed analysis of the experimental values of $\Delta_2$ for various Co(II) compounds may be found in Ref.~\onlinecite{doi:10.1142/S0217979221300061}. All of these materials are expected to be well described by a \mbox{$J = \frac{1}{2}$} model, which can be verified experimentally by the observation of the well-defined local $j_{1/2} \to j_{3/2}$ excitations in an energy range larger than the dispersion of magnetic excitations \cite{PhysRevB.98.024415,ross2017single,songvilay2020kitaev,kim2021antiferromagnetic}. For the materials of interests, the trigonal splitting is therefore not sufficiently large to invalidate the $j_{1/2}$ picture, but may nonetheless have a considerable effect on the magnetic couplings, as discussed in the following sections.

\subsection{Survey of Hopping in Real Materials}
\label{sec:hops}

As noted in previous sections, the key question for the realization of dominant Kitaev coupling in high-spin $d^7$ compounds is the relative importance of direct metal-metal vs. ligand-assisted hopping processes. In terms of the usual Slater-Koster hoppings, the downfolded $d$-only hoppings are given approximately by:
\begin{align}\label{eqn:t1}
t_1 \approx & \ \frac{1}{2} t_{dd}^\pi \\
t_2 \approx & \ -\frac{1}{2} t_{dd}^\pi + \frac{(t_{pd}^\pi)^2}{\Delta_{pd}} \\
t_3 \approx & \ \frac{3}{4}t_{dd}^\sigma\\
t_4 \approx & \ -\frac{1}{2} t_{dd}^\sigma \\
t_5 \approx & \ t_{dd}^\pi \\
t_6 \approx & \ \frac{\sqrt{3}}{4} t_{dd}^\sigma - \frac{t_{pd}^\pi t_{pd}^\sigma}{\Delta_{pd}}
\label{eqn:t6}
\end{align}
where $\Delta_{pd} $ is the charge-transfer energy from Co $d$ to the ligand $p$ orbitals. Here, we have ignored contributions from $t_{dd}^\delta$. For $t_2$ and $t_6$, the second terms represent ligand-assisted hopping. There are two main factors that can affect these: (i) the Co-Co bond lengths (which primarily affect the direct overlap of metal $d$-orbitals on adjacent metal sites) and (ii) the degree of hybridization with the ligands (as quantified by the relative weight of the $p$-orbitals in the $d$-orbital Wannier functions, which scales as $\propto t_{pd}/\Delta_{pd}$). 

\begin{figure}[t]
\includegraphics[width=\linewidth]{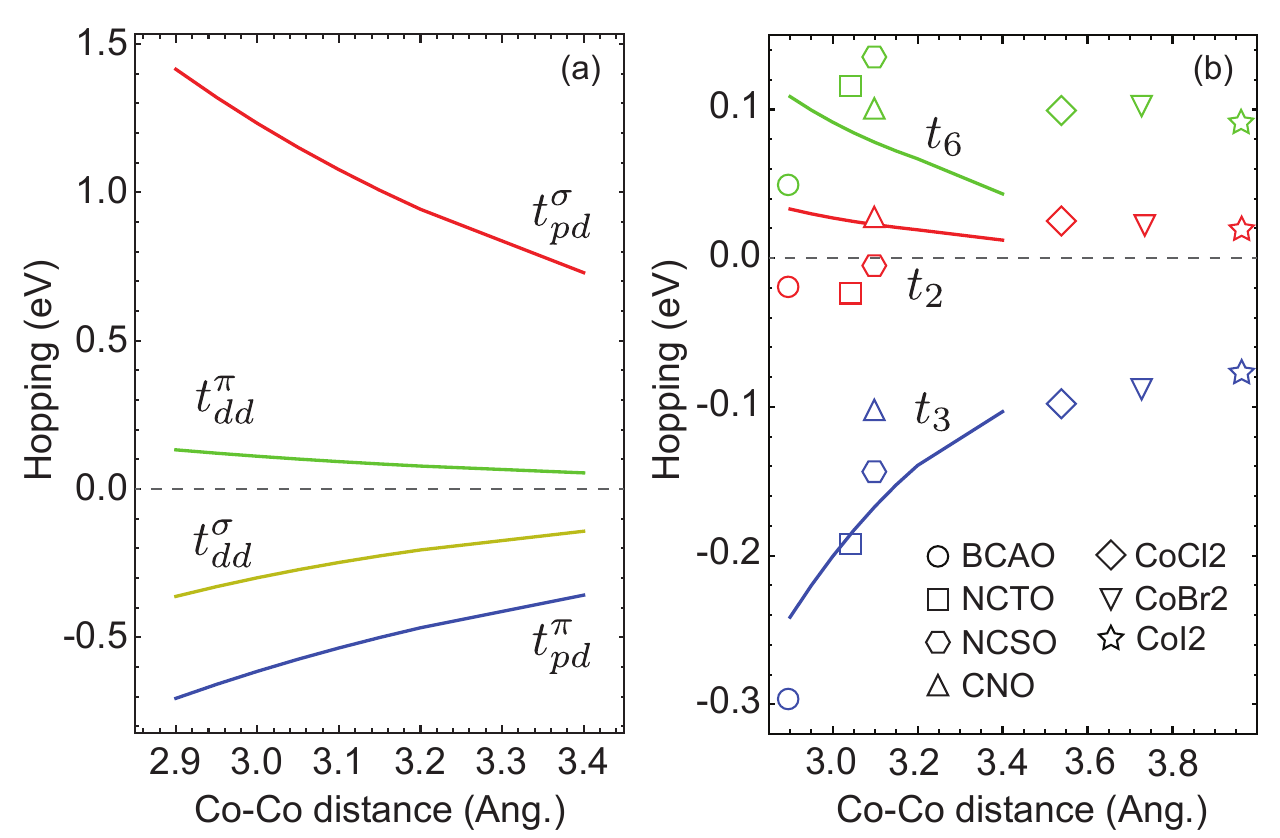}
\caption{Evolution of the relevant hoppings as a function of Co-Co distance. (a) Hoppings in the full $(3d+2p)$ scheme for hypothetical symmetrically stretched structures of CoO (see text). (b) Hoppings in the downfolded $d$-only scheme. Solid lines correspond to hypothetical stretched cubic CoO. Points correspond to real materials; BCAO = BaCo$_2$(AsO$_4$)$_2$, NCTO = Na$_2$Co$_2$TeO$_6$, NCSO = Na$_3$Co$_2$SbO$_6$, CNO = CoNb$_2$O$_6$. }
\label{fig:hops_var}
\end{figure}

In order to investigate the hopping trends with bond-length dependence, we first considered hypothetical cubic CoO (NaCl type; $Fm\bar{3}m$) structures with a symmetrically stretched unit cells, allowing us to investigate a continuous range of geometries. This construction maintains 90$^\circ$ Co-O-Co bond angles, which deviates slightly from real materials, but allows us to consider ideal trends. In order to estimate the hoppings, we employed fully relativistic density functional theory calculations performed with FPLO\cite{koepernik1999full,opahle1999full} at the GGA (PBE\cite{perdew1996generalized}) level.  Hopping integrals were extracted by formulating Wannier orbitals via projection onto atomic $p$ and/or $d$-orbitals. For example, the Slater-Koster hoppings were extracted from the CoO calculations by fitting with an 8-band $(3d+2p)$ model including explicitly the O orbitals. Results are shown in Fig.~\ref{fig:hops_var}(a) as a function of Co-Co distance. 

As expected, the largest magnitude corresponds to $t_{pd}^\sigma$, which quantifies the hopping between ligand $p$ and metal $e_g$ orbitals. This is followed in magnitude by $t_{pd}^\pi$, which quantifies the hopping between ligand $p$ and metal $t_{2g}$ orbitals. The difference in magnitude between $t_{pd}^\sigma$ and $t_{pd}^\pi$ results in a larger degree of metal-ligand hybridization for the Wannier $e_g$ orbitals than the $t_{2g}$ orbitals. This is, of course, the origin of octahedral crystal field splitting $\Delta_1$, as described by standard ligand field theory. From the signs of the Slater-Koster hoppings, one can see that the contribution to $t_2$ and $t_6$ from ligand-assisted hopping is always positive, while the direct hopping contribution is always negative. The balance of this competition is therefore governed by the charge-transfer energy $\Delta_{pd}$. For $3d$ transition metal compounds, this quantity is typically large, which reduces ligand-metal hybridization relative to their $4d$ and $5d$ counterparts. It is precisely this effect that leads to smaller $t_{2g}-e_{g}$ splitting $\Delta_1$ for $3d$ transition metal compounds \cite{gray1965electrons}, which is a key requirement for stability of the high-spin state in Co $3d^7$ compounds. For this reason, ligand-assisted hopping is expected to be suppressed overall. This leads to strong competition between hopping processes and overall suppression of $t_2$ and $t_6$. In order to demonstrate this effect, we show in Fig.~\ref{fig:hops_var}(b) (solid lines) the {\it downfolded} hoppings extracted for the hypothetical CoO structures by Wannier fitting with a 5-band (3d-only) basis. The results are compatible with (but not exactly given by) Eq'ns (\ref{eqn:t1}) - (\ref{eqn:t6}) with $\Delta_{pd} \sim 4.5$ eV. For all lattice constants, we find that $t_3$ is the dominant hopping rather than the ligand-assisted $t_2$ and $t_6$.

While we leave complete discussion of individual materials for later work, it is useful to confirm realistic ranges of hoppings. In order to do so, we performed additional calculations on several prominent oxide materials based on literature structures: CoNb$_2$O$_6$ (Ref.~\onlinecite{sarvezuk2011new}), BaCo$_2$(AsO$_4$)$_2$ (Ref.~\onlinecite{djordevic2008baco2}), Na$_3$Co$_2$SbO$_6$ (Ref.~\onlinecite{songvilay2020kitaev}), and Na$_2$Co$_2$TeO$_6$ (Ref.~\onlinecite{xiao2019crystal})\footnote{The Na$_2$Co$_2$TeO$_6$ structure contains disorder in the Na position, in which each Na position has occupancy 2/3. To perform calculations, we artificially increased the occupancy to 1, which corresponds to Na$_3$Co$_2$TeO$_6$. It is expected this change in the filling should have minimal impact on the computed hoppings.}, as well as triangular lattice compounds CoCl$_2$, CoBr$_2$ and CoI$_2$ according to crystal structures from Ref.~\onlinecite{wilkinson1959neutron,wyckoff1963crystal}. These edge-sharing Co(II) compounds span a wide range Co-Co distances, e.g. from $\sim 2.9$ \AA \ in BaCo$_2$(AsO$_4$)$_2$\cite{djordevic2008baco2} to $\sim 3.9$ \AA \ in CoI$_2$\cite{wyckoff1963crystal}. For each case, we employed fully relativistic density functional theory calculations performed with FPLO\cite{koepernik1999full,opahle1999full} at the GGA (PBE\cite{perdew1996generalized}) level (as above). Hoppings in the downfolded $d$-only basis were estimated by projection onto $d$-orbitals, for which the cubic projection coordinates were defined to be orthogonal but minimize the difference with the corresponding Co-X (X = O, Cl, Br, I) bond vectors in the (distorted) octahedra.

Computed hoppings for the real materials are shown in Fig.~\ref{fig:hops_var}(b). In general, $t_1$, $t_4$ and $t_5$ are small, and are thus omitted from the figure. From these results, it is evident that the physical region corresponds to large direct hopping $t_3$ and subdominant ligand-mediated hopping $t_6$. Similarly, the ligand-mediated $t_2$ is significantly suppressed, such that $|t_2| \sim |t_1|,|t_4|,|t_5| \lesssim 0.05$ eV. We find that this applies equally to Co oxides and halides. A similar situation\cite{wellm2021frustration} was recently proposed for Na$_2$BaCo(PO$_4$)$_2$. On this basis, we conclude: for edge-sharing materials with Co-Co bond lengths $\sim$ 3.0 \AA, direct hopping almost certainly dominates. This differs from the previous theoretical works predicting large Kitaev couplings\cite{liu2018pseudospin,doi:10.1142/S0217979221300061,liu2020kitaev,sano2018kitaev}, which considered ligand-mediated hopping $t_2$ and $t_6$ to be the largest. This discrepancy calls for a reexamination of the magnetic couplings.

\section{Magnetic Couplings}
\subsection{General Form}
For ideal edge-sharing bonds with $C_{2v}$ symmetry, the magnetic couplings may be written in the familiar form\cite{rau2014generic}:
\begin{align}
\mathcal{H}_{ij} =& \  J \ \mathbf{S}_i \cdot\mathbf{S}_j + K \ S_i^\gamma S_j^\gamma 
+ \Gamma \left(S_i^\alpha S_j^\beta + S_i^\beta S_j^\alpha\right)
\nonumber \\
& + \Gamma^\prime \left(S_i^\alpha S_j^\gamma + S_i^\gamma S_j^\alpha +S_i^\beta S_j^\gamma + S_i^\gamma S_j^\beta\right)
\end{align}
where $(\alpha,\beta,\gamma) = (x,y,z)$ for the Z-bonds, $(y,z,x)$ for the X-bonds, and $(z,x,y)$ for the Y-bonds, in terms of the global $xyz$ coordinates. This definition turns out to be convenient for the case of trigonal splitting $\Delta_2 = 0$.  For finite $\Delta_2$, following Ref.~\onlinecite{lines1963magnetic}, the alterations to the nature of the local moments are expected to induce significant uniaxial anisotropy along the trigonal axis.
This axial anisotropy is more apparent in alternative local XXZ coordinates shown in Fig.~\ref{fig:hops}. In particular, for each bond we define local coordinates: $\hat{e}_1$ is parallel to the bond and $\hat{e}_3 = (\hat{x}+\hat{y}+\hat{z})/\sqrt{3}$ is along the global trigonal axis. Thus, the couplings may be written \cite{ross2011quantum,maksimov2019anisotropic}:
\begin{align}
\mathcal{H}_{ij} = & \  J_{xy} \left( S_i^1 S_j^1 + S_i^2 S_j^2\right) +J_z S_i^3 S_j^3 
 \\ \nonumber
& \ 
+2J_{\pm\pm} \left(S_i^1S_j^1 - S_i^2 S_j^2 \right) + J_{z\pm} \left(S_i^3 S_j^2 + S_i^2 S_j^3\right)
\end{align}
where the superscript numbers refer to the local directions. 
The two parameterizations may be related\cite{maksimov2019anisotropic} via:
\begin{align}
J_{xy} = & \ J + \frac{1}{3}\left( K - \Gamma - 2\Gamma^\prime\right) \label{eqn:Jxy}\\
J_z = & \ J + \frac{1}{3}\left(K + 2\Gamma + 4\Gamma^\prime\right)\label{eqn:Jz}\\
J_{\pm\pm} = & \ -\frac{1}{6} \left( K + 2\Gamma - 2\Gamma^\prime\right)\\
J_{z\pm} = & \ -\frac{\sqrt{2}}{3}\left( K - \Gamma + \Gamma^\prime\right)
\end{align}
A similar parameterization was also employed in Ref.~\onlinecite{doi:10.1142/S0217979221300061,liu2020kitaev}. Both parameterizations are employed below, where appropriate.

\subsection{Method and Downfolding Schemes}

\begin{figure}[t]
\includegraphics[width=\linewidth]{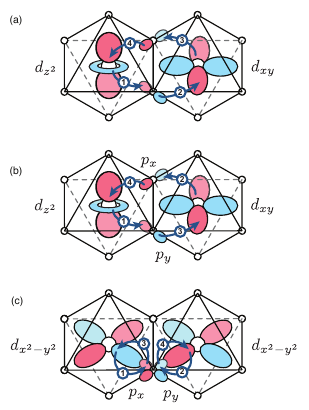}
\caption{Example hopping paths for holes contributing to magnetic exchange. Terms resulting from (a) and (b) are captured in the downfolded $d$-only scheme. Terms resulting from (c) are not captured, and must be estimated from perturbative expressions (see text). }
\label{fig:exchangepaths}
\end{figure}

In previous works, the magnetic couplings have been discussed in terms of perturbation theory in both the full $d+p$ basis, and the downfolded $d$-only basis. It is useful to discuss how terms in each scheme are related. 

As an example of this downfolding, we may consider the ligand-mediated exchange process shown in Fig.~\ref{fig:exchangepaths}(a), in which a hole hops from a metal $d$-orbital to a ligand $p$-orbital, and subsequently to a $d$-orbital on an adjacent metal site, thus yielding an excited $(d^8,d^6)$ configuration. From this excited configuration, the reverse process returns the system to the ground $(d^7,d^7)$ configuration. Such processes contribute terms of order $\mathcal{O}[  \ (t_{pd}^2/\Delta_{pd})^2/U_{dd} \ ]$ to the magnetic exchange, where $\Delta_{pd}$ is the charge transfer energy, and $U_{dd}$ is the Coulomb repulsion between $d$-electrons. The same process can also be captured within a $d$-only picture, by identifying $t_{\rm eff} = t_{pd}^2/\Delta_{pd}$ as an effective ligand-mediated $d-d$ hopping, as discussed above. In this case, the magnetic couplings arise at order $\mathcal{O}[ \ t_{\rm eff}^2 / U_{dd} \ ]$. Such terms are naturally captured in the downfolded $d$-only scheme.

We next consider the ``cyclic'' exchange process depicted in Fig.~\ref{fig:exchangepaths}(b). In this case, a hole first hops from one metal to a ligand, and then a second hole hops from the second metal to another ligand. The system is returned to the ground $(d^7, d^7)$ configuration by two additional hops of the holes from the ligands to opposite metal sites from where they started. Such processes contribute to the magnetic exchange at order $\mathcal{O}[ \ (t_{pd}^2/\Delta_{pd})^2/(2\Delta_{pd}) \ ]$. In the downfolded $d$-only basis, these processes are not distinguished from those in Fig.~\ref{fig:exchangepaths}(a). Rather, they reflect the fact that the Coulomb repulsion within the $d$-orbital Wannier functions is {\it renormalized} to $\tilde{U}^{-1} \sim U_{dd}^{-1} + (2\Delta_{pd})^{-1}$ via hybridization of the $d$- and $p$-orbitals. Thus both the ``antiferromagnetic'' superexchange and cyclic exchange processes are captured, in principle, by exchange contributions at order $\mathcal{O} [ \ t_{\rm eff}^2/ \tilde{U} \ ]$, assuming the hoppings and Coulomb repulsion terms are suitably renormalized.  

Finally, we consider the process depicted in Fig.~\ref{fig:exchangepaths}(c). In this case, two holes hop from their parent metal sites to the same ligand, but occupy different orbitals. They interact with eachother via Hund's coupling at the ligand site, and then hop back to their parent metal sites. Such processes constitute the usual Goodenough-Kanamori ferromagnetic superexchange. When downfolded into the $d$-only basis, these processes reflect the fact that the tails of the Wannier orbitals extend onto the ligands, which results in an enhancement of the effective {\it intersite} Hund's coupling. Such terms are not explicitly included in our $d$-only model Eq'n (\ref{eqn:overallham}). We therefore estimate the effects of the additional contributions using previously derived perturbative expressions. From Ref.~\onlinecite{liu2018pseudospin}, there is a correction to both $J$ and $K$ given approximately by:
\begin{align}
\delta J \approx & \ -\frac{\gamma}{\Delta_{pd}^2} \left( \frac{5}{2}(t_{pd}^\sigma)^4+\frac{3}{2}(t_{pd}^\sigma)^2 (t_{pd}^\pi)^2 +(t_{pd}^\pi)^4  \right) \label{eqn:deltaJ}
\\
\delta K \approx & \ -\frac{\gamma}{\Delta_{pd}^2} \left(\frac{1}{2}(t_{pd}^\sigma)^2 (t_{pd}^\pi)^2-\frac{1}{2}(t_{pd}^\pi)^4  \right) 
\\
\gamma =& \  \frac{40J_H^p}{81(\Delta_{pd}+U_p/2)^2}
\end{align}
where $J_H^p$ is the Hund's coupling at the ligand, $U_p$ is the excess Coulomb repulsion at the ligand ion, $\Delta_{pd} \approx 4.5$ eV is the charge-transfer gap. We take the same approximations as Ref.~\onlinecite{liu2018pseudospin} ($U_p = 0.7 \ U_{t2g}, \ J_H^p = 0.3 \ U_p$), and consider $t_{pd}^\sigma \approx 1$ eV, $t_{pd}^\pi \approx -0.5$ eV, according to Fig.~\ref{fig:hops_var}(a). From this, we estimate the correction to the Kitaev coupling to be negligible $\delta K \sim 0.1$ meV, while the corrections to the Heisenberg coupling may be typically in the range $\delta J \sim -2$ to $-6$ meV.  The reason for the discrepancy in magnitudes between $\delta J$ and $\delta K$ is that the dominant contribution to $\delta J$ comes from hopping between the $e_g$ and $p$ orbitals (depicted in Fig.~\ref{fig:exchangepaths}(c)), as parameterized by the large hopping $t_{pd}^\sigma$. This process does not contribute to $K$. The small correction $\delta K$ is henceforth neglected. 

In the following, we compute all exchange contributions that are captured within the downfolded $d$-only basis using exact diagonalization of the model Eq'n (\ref{eqn:overallham}) for two Co sites. The couplings are extracted by projecting onto the ideal $j_{1/2}$ doublets defined in eq'n (\ref{eqn:j12a}, \ref{eqn:j12b}). This procedure is described in more detail in Ref.~\onlinecite{winter2016challenges}, and is guaranteed to yield couplings that converge to the results of perturbation theory with respect to $\mathcal{H}_{\rm hop}$ for small values of $t_n/U$ (with all other contributions to the Hamiltonian, which appear in $\mathcal{H}_i$, being treated exactly). The reason for adopting this method is that analytical perturbation theory is generally challenging for finite $\Delta_2$ and the full set of allowed hoppings $t_{1-6}$. The expressions below are intended to be used with hopping integrals extracted from DFT, which are assumed to be suitably renormalized. The correction $\delta J$ must be added to the computed couplings to estimate the omitted ligand exchange processes depicted in Fig.~\ref{fig:exchangepaths}(c). 

\begin{figure*}[t]
\includegraphics[width=\linewidth]{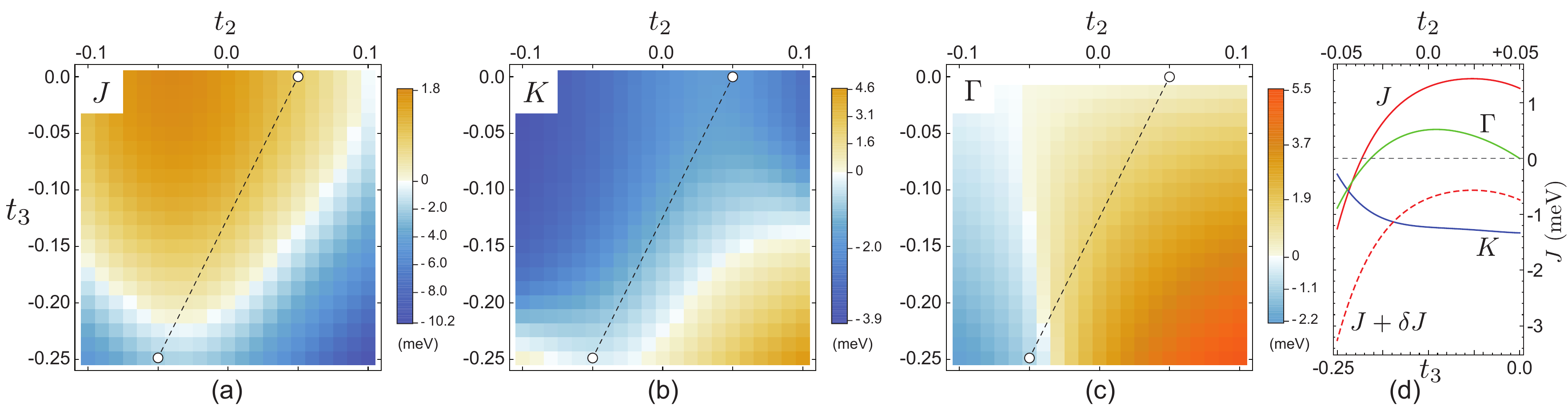}
\caption{(a-c) Evolution of the magnetic couplings in the $d$-only model for ideal edge-sharing bond with no trigonal distortion ($\Delta_2 =0$) and $\Delta_1 = 1.1$ eV, $U = 3.25$ eV, $J_H = 0.7$ eV,  $t_1 = |t_3|/4, t_4 =t_5= -|t_3|/4, t_6 = +0.1$ eV. The ferromagnetic correction $\delta J$ due to ligand exchange processes is not included (see text). (d) Computed couplings along the path indicated by a dashed line in (a-c), interpolating between the direct and ligand-assisted hopping regimes. A correction $\delta J  = -2$ meV has been added. The small correction $\delta K$ has been neglected.}
\label{fig:kcouplings}
\end{figure*}

\subsection{General Hopping Dependence}  \label{sec:geninteractions}
In this section, we first consider the magnetic couplings for the case $\Delta_2 = 0$, for which $\Gamma^\prime = 0$ strictly. Up to second order in $d-d$ hopping (and fourth order in $d-p$ hopping), the couplings may be written:
\begin{align}
J = & \ \mathbf{t} \cdot \mathbb{M}_J \cdot \mathbf{t}^T+\delta J \\
K =& \  \mathbf{t} \cdot \mathbb{M}_K \cdot \mathbf{t}^T\\
\Gamma =& \ \mathbf{t} \cdot \mathbb{M}_\Gamma \cdot \mathbf{t}^T
\end{align}
where:
\begin{align}
\mathbf{t} = \left( t_1 \ t_2 \ t_3 \ t_4 \ t_5 \ t_6 \right)
\end{align}
and $\mathbb{M}$ parameterizes the contributions that are captured by (downfolded) $d-d$ hopping. The matrices are a function of $F_n, \lambda, \Delta_n$. To estimate $\mathbb{M}$, we computed the magnetic couplings for a grid of hoppings $-0.05 < t_n < +0.05$ eV using exact diagonalization of the full $d$-orbital model $\mathcal{H}_U+\mathcal{H}_{\rm CFS}+\mathcal{H}_{\rm SOC} + \mathcal{H}_{\rm hop}$ for two metal sites, and fit the resulting couplings. 
For this purpose, we use $\Delta_1 = 1.1$ eV and $\lambda = 0.06$ eV, which is consistent with estimates from DFT in the previous sections and $U_{t2g} = 3.25$ eV, and $J_{t2g} = 0.7$ eV, following Ref.~\onlinecite{das2021xy}. This provides an estimate of the couplings in the perturbative regime:
\begin{align}
\mathbb{M}_J = & \ \left(
\begin{array}{c|cccccc}
& t_1 & t_2 & t_3 & t_4 & t_5 &t_6 \\
\hline
t_1 & -55 & 0 & 143 & 10 & 2 & 0 \\
t_2 & 0 & -76 & 0 & 0 & 0 & -77 \\
t_3 & 0 & 0 & -33 & 2 & 2 & 0 \\
t_4 & 0 & 0 & 0 & 260 & 0 & 0 \\
t_5 & 0 & 0 & 0 & 0 & 259 & 0 \\
t_6 & 0 & 0 & 0 & 0 & 0 & 165 
\end{array}
 \right)
\end{align}
\begin{align}
\mathbb{M}_K = & \ \left(
\begin{array}{c|cccccc}
& t_1 & t_2 & t_3 & t_4 & t_5 &t_6 \\
\hline
t_1 & 128 & 0 & -119 & -9 & 35& 0 \\
t_2 & 0 & -108 & 0 & 0 & 0 & 86 \\
t_3 & 0 & 0 & -8 & -2 & 5 & 0 \\
t_4 & 0 & 0 & 0 & -4 & 0 & 0 \\
t_5 & 0 & 0 & 0 & 0 & 1 & 0 \\
t_6 & 0 & 0 & 0 & 0 & 0 & -147 
\end{array}
 \right)
\end{align}
\begin{align}
\mathbb{M}_\Gamma = & \ \left(
\begin{array}{c|cccccc}
& t_1 & t_2 & t_3 & t_4 & t_5 &t_6 \\
\hline
t_1 & 0 & -34 & 0 & 0 & 0 & 49 \\
t_2 &0 & 0 & -116 & -2 & 1 & 0 \\
t_3 &0 & 0 & 0 & 0 & 0 & -31 \\
t_4 &0 & 0 & 0 & 0 & 0 & -67 \\
t_5 &0 & 0 & 0 & 0 & 0 & 0 \\
t_6 &0 & 0 & 0& 0 & 0 & 0 
\end{array}
 \right)
\end{align}
in units of $10^{-3}$/eV. Recall, for real materials we generally anticipate $|t_3| > |t_6| > |t_1| \sim |t_2| \sim |t_4| \sim |t_5|$. Furthermore, $t_1>0$, $t_3 <0$, $t_4 < 0$, $t_6>0$ (see section \ref{sec:hops}). These results highlight several key aspects:

 {\it Heisenberg $J$}: For $J$, there are various contributions of different sign. Those arising from hopping between $t_{2g}$ orbitals ($t_1, t_2, t_3$) are exclusively ferromagnetic. Processes involving hopping between $e_g$ orbitals ($t_4,t_5$) are exclusively antiferromagnetic. The terms related to $e_g$-$t_{2g}$ hopping tend to be antiferromagnetic $\propto t_6^2$ and $t_2t_6$ given that {\it ab-initio} tends to yield $t_2 <0$ and $t_6 >0$.

{\it Kitaev $K$}: There are also different contributions to $K$ of varying sign. Hopping between $e_g$ orbitals ($t_4,t_5$) makes little contribution to the anisotropic couplings overall. The sign of the contribution from $t_{2g}$-$t_{2g}$ hopping depends on the balance of transfer integrals: terms $\propto t_2^2$ are ferromagnetic, while terms $\propto t_1^2$ and $t_1t_3$ are antiferromagnetic. Contributions related to $t_{2g}$-$e_g$ hopping may take both signs: terms $\propto t_6^2$ are ferromagnetic, while terms $\propto t_2t_6$ depend on the sign of $t_2$. 

{\it Off-diagonal $\Gamma$}: For the off-diagonal couplings, the primary contribution arises at order $t_2 t_3$, and as a result $\text{sign}(\Gamma)$ is typically the same as $-\text{sign}(t_2t_3)$. There are no contributions that are diagonal with respect to the hopping pairs. A similar result appears in Ref.~\onlinecite{liu2018pseudospin}. 

The appearance of hopping combinations such as $t_2t_6$, which do not conserve $t_{2g}$ and $e_g$ occupancy, may appear surprising at first. If the ground state doublets have approximate configuration $(t_{2g})^5 (e_g)^2$, one might expect terms mixing the occupancy to be forbidden at low orders, because they do not connect ground states. However, in reality, neither occupancy is preserved by either spin-orbit coupling or the full Coulomb terms, which are treated exactly (not perturbatively) in this approach. 

For general parameters, we expect all three couplings to be finite. The computed hopping-dependence of $K, J, \Gamma$ are shown in Fig.~\ref{fig:kcouplings} for the choice $t_1 = |t_3|/4, t_4 =t_5= -|t_3|/4, t_6 = +0.1$ eV. These ratios of hoppings are compatible with the {\it ab-initio} estimates in section \ref{sec:hops}. With this choice, we interpolate between the limits of dominant ligand-mediated vs. direct hopping. 

In the hypothetical regime of pure ligand-mediated hopping (finite $t_2,t_6>0$ ), we find $\Gamma = 0$. If we ignore the ferromagnetic correction $\delta J$, then $J > 0$ is antiferromagnetic and $K < 0$ is ferromagnetic. These findings verify expectations from perturbation theory for this limit\cite{liu2018pseudospin,sano2018kitaev}. The Kitaev coupling is the largest, with values $|K/J| \sim 1 - 10$ depending on the precise balance of hoppings. If we consider also the ferromagnetic correction $\delta J \sim -2$ to $-6$ meV discussed in the previous section, the overall sign of $J$ should reverse, and the magnitude may be suppressed, such that dominant Kitaev coupling is possible with some tuning.

By contrast, for the physically relevant region of large $t_3$ and finite values of all hoppings, we anticipate that ferromagnetic $J < 0$ is the dominant coupling, particularly due to contributions $\propto t_1 t_3$ and the correction $\delta J$. In fact, $\delta J$ (which is just the regular ferromagnetic super-exchange for 90$^\circ$ bonds\cite{goodenough1963magnetism}) is the largest contribution. All possible combinations of signs of $K$ and $\Gamma$ are possible depending on the balance of hoppings, but their magnitude is suppressed relative to $J$. For very short Co-Co bond lengths, where the direct hopping contribution to $t_2$ is the largest ($t_2 < 0$), the tendency is for $K, \Gamma <0$. For longer bond lengths, where ligand-mediated contributions are the largest ($t_2 >0$), then $K,\Gamma > 0$.

\subsection{Effect of Trigonal Distortion}  \label{sec:trig}

We next consider the effects of trigonal distortion. Given the relatively small value of the atomic SOC constant $\lambda_{\rm Co} \approx 60$ meV, small distortions may be relevant for Co(II) compounds. This make clarifying the size and sign of $\Delta_2$ important for modelling such materials.

 In general, for $\Delta_2 < 0$, as the moments become more axial, components of the exchange along the $\hat{e}_3$ direction are expected to be enhanced compared to the $\hat{e}_1$ and $\hat{e}_2$ directions\cite{lines1963magnetic}. As a result $J_{xy}$ and $J_{\pm \pm}$ should be suppressed relative to $J_z$ and $J_{z\pm}$. Trigonal elongation $\Delta_2 > 0$ should have the opposite effect. As the moments become more planar, $J_{xy}$ and $J_{\pm \pm}$ should be relatively enhanced. 

\begin{figure}[t]
\includegraphics[width=0.8\linewidth]{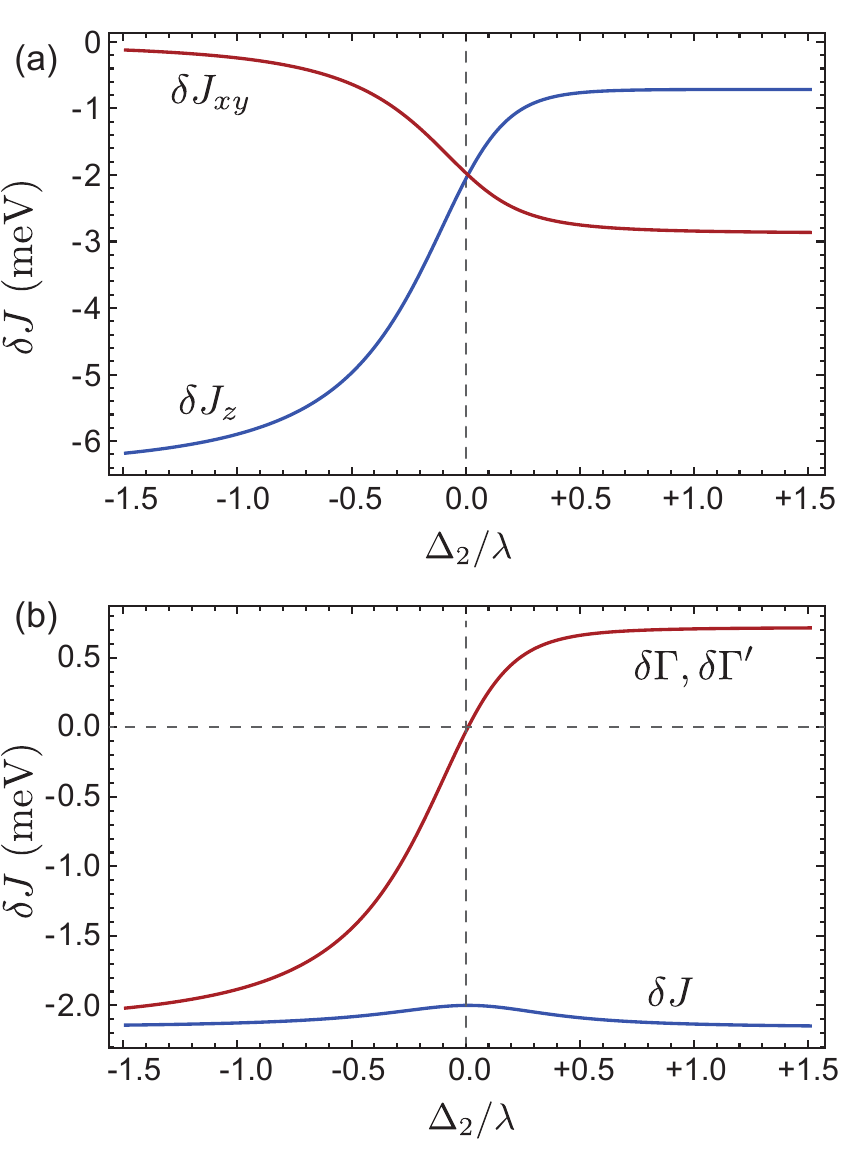}
\caption{Anisotropic ferromagnetic corrections as a function of trigonal distortion, for $\delta J_0 = -2$ meV. (a) Local XXZ scheme. (b) Global $J,K,\Gamma,\Gamma^\prime$ scheme. }
\label{fig:trigcoup}
\end{figure}

\begin{figure*}[t]
\includegraphics[width=\linewidth]{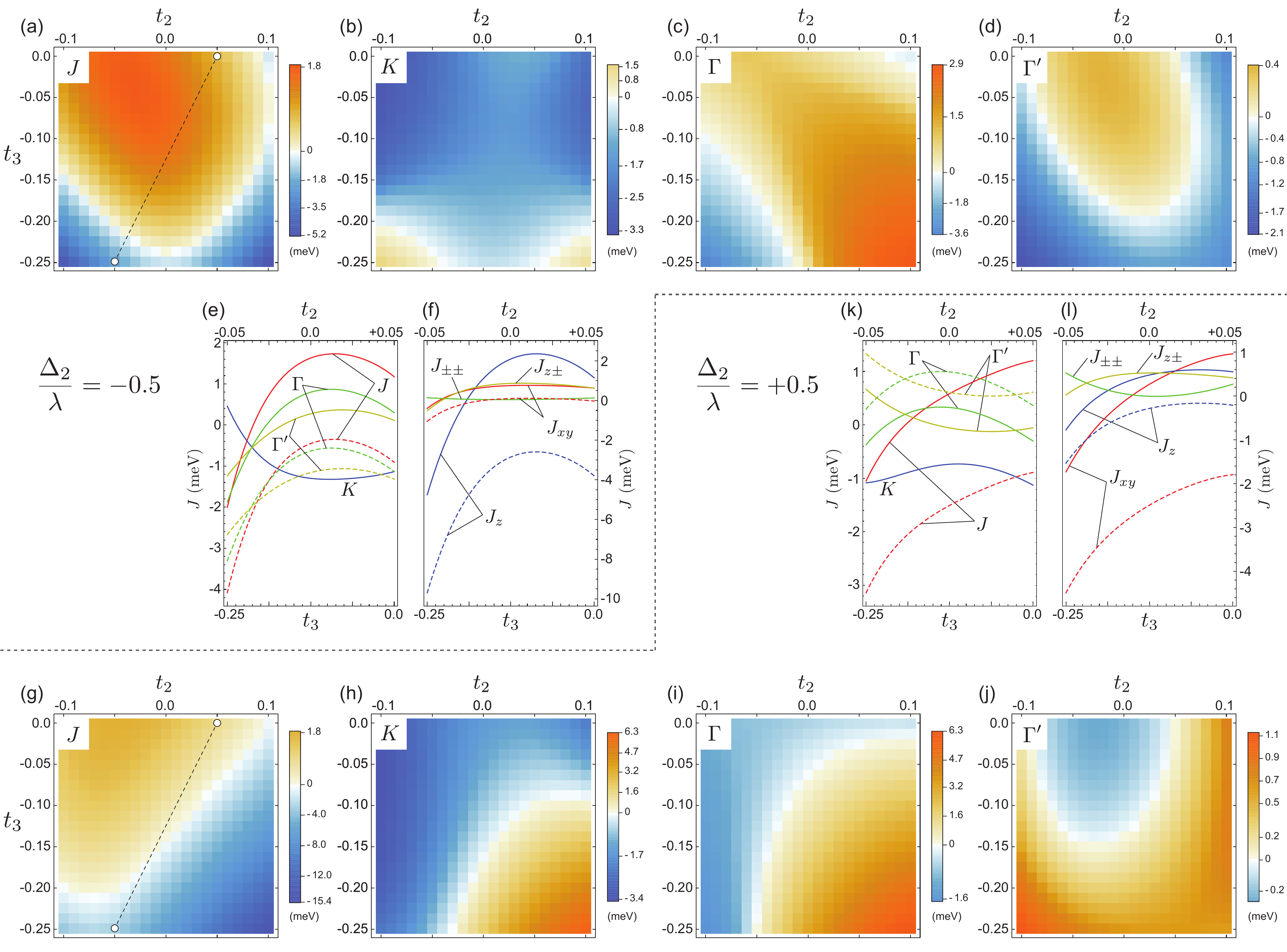}
\caption{Magnetic couplings for ideal edge-sharing bond with finite trigonal distortion. The parameters are otherwise the same as Fig.~\ref{fig:kcouplings}. (a-f): $\Delta_2/\lambda = -0.5$, (g-l): $\Delta_2/\lambda = +0.5$. The ferromagnetic corrections $\delta J,\delta\Gamma, \delta \Gamma^\prime$ due to ligand exchange processes are not included (see text).
(e,k): Couplings along the path interpolating between direct and ligand-assisted hoppings depicted in (a,g) in the $J,K,\Gamma,\Gamma^\prime$ scheme. (f,l): Couplings along the path in the $J_z,J_{xy},J_{z\pm},J_{\pm\pm}$ scheme. For (e,f,k,l), solid lines indicate the results of $d$-$d$ exchange only, and dashed lines indicate corrected values $J+\delta J$, according to $\delta J_0 = -2$ meV.}
\label{fig:trigcouplings}
\end{figure*}

Following Ref.~\onlinecite{lines1963magnetic,liu2020kitaev}, the ferromagnetic corrections to $J$ resulting from ligand exchange processes are rendered anisotropic, with:
\begin{align}
\delta J_{xy} \approx & \ u_{xy}^2 \ \delta J_0 \\
\delta J_{z} \approx & \ u_{z}^2 \ \delta J_0 \\
u_{xy} = & \ \frac{3}{5}\left(2\sqrt{3}c_1 c_3 + 2c_2^2\right) \\
u_{z} =& \  \frac{3}{5}\left(1+2(c_1^2-c_3^2)\right)
\end{align}
with $c_n$ given in Fig.~\ref{fig:single}(b) and $\delta J_0$ defined according to eq'n (\ref{eqn:deltaJ}). Thus, in the $J,K,\Gamma,\Gamma^\prime$ parameterisation, these corrections correspond to:
\begin{align}
\delta J =& \  \frac{1}{3} \left(  u_z^2+2u_{xy}^2 \right) \delta J_0 \\
\delta \Gamma = & \ \delta \Gamma^\prime  = \frac{1}{3} \left( u_z^2 - u_{xy}^2\right) \delta J_0
\end{align}
Here, we have continued to neglect the small corrections $\delta K_0$. 
As with the undistorted case, the corrections to the anisotropic couplings $J_{\pm\pm}$ and $J_{z\pm}$ are predicted to be small. They have also been neglected in the following. The evolution of the ferromagnetic corrections with distortion are shown in Fig.~\ref{fig:trigcoup}.
For $\Delta_2 < 0$, the ratio $\delta J_z / \delta J_{xy}> 1$ is, in principle, unbounded (and should increase continuously with trigonal distortion). For $\Delta_2 > 0$ the degree of anisotropy is restricted, because the distortion-induced effects are bounded $1/4 \lesssim J_z/J_{xy} <1$. The lower bound is reached for large $\Delta_2$, where the orbital moment is quenched, thus restoring the full degeneracy of the $S = 3/2$ states. As discussed above, a low-energy model including only the lowest doublet would no longer be sufficient, so this limit does not represent a physically sensible model. In terms of global cubic coordinates [Fig.~\ref{fig:trigcoup}(b)], the trigonal distortion primarily introduces off-diagonal couplings, where $\Delta_2 <0$ tends to be associated with $\delta \Gamma,\delta \Gamma^\prime < 0$, and vice versa.

To explore the exchange contributions from (downfolded) $d$-$d$ hopping, we recomputed the couplings using exact diagonalization 
with significant distortion $\Delta_2/\lambda = \pm 0.5$ to emphasize the effects. Results are shown in Fig.~\ref{fig:trigcouplings} for the choice $t_1 = |t_3|/4, t_4 =t_5= -|t_3|/4, t_6 = +0.1$ eV, which is compatible with the {\it ab-initio} estimates. In Fig.~\ref{fig:trigcouplings}(e,f) and (k,l), we also show the effect of corrections $\delta J$. The results are as follow:

{\it Trigonal compression}: For $\Delta_2 < 0$, as shown in Fig.~\ref{fig:trigcouplings}(a-e), we find all four of the couplings $J,K,\Gamma,\Gamma^\prime$ may be of similar magnitude. This is particularly true in the region of large ligand-assisted hopping. For the physically relevant region of large direct hopping ($t_3 \gg t_2$), we find that $K$ is still relatively suppressed (same as for $\Delta_2 = 0$), but large $\Gamma,\Gamma^\prime$ are induced, with $\text{sign}(\Gamma, \Gamma^\prime)$ typically being the same as $\text{sign}(J)$. These results are more easily interpreted in the alternative XXZ parameterization shown in Fig.~\ref{fig:trigcouplings}(f). In particular, as the local moments become more axial with larger trigonal distortion, the coupling becomes dominated by a ferromagnetic Ising exchange $J_z$. Overall, the estimated ferromagnetic correction $\delta J_z$ is quite large compared to the regular $d$-$d$ contributions. 
For the physically relevant region, we anticipate $J_{xy}=-2$ to $0$ meV, $J_z = -3$ to $-10$ meV, $J_{\pm\pm} = -0.5$ to $+0.5$ meV, and $J_{z\pm} = -0.5$ to $+1.5$ meV for significant trigonal distortion of $\Delta = -\lambda/2$. . As a result, we expect such materials to be described mostly by Ising couplings with a common axis for every bond. 

{\it Trigonal elongation}: For $\Delta_2 > 0$, we find that $K$ is less suppressed. The distortions induce off-diagonal couplings following roughly $\text{sign}(\Gamma, \Gamma^\prime)$ typically the same as $-\text{sign}(J)$. In the XXZ parameterization, this corresponds to an enhancement of $J_{xy}$. In the hypothetical ligand-assisted hopping region, we find that $J_z$ may be almost completely suppressed due to different values of the ferromagnetic shifts $\delta J_z$ and $\delta J_{xy}$. While $J_{xy}$ appears to be the largest coupling in this limit, the bond-dependent couplings $J_{z\pm}$ and $J_{\pm \pm}$ may also remain significant. For the physically relevant region, we find that $J_{xy}$ is typically the dominant coupling, with $J_{xy}/J_z\sim 4$, which is the hypothetical limit. Overall, we anticipate $J_{xy}=-2$ to $-10$ meV, $J_z = -0.5$ to $-4$ meV, $J_{\pm\pm} = -2$ to $+1$ meV, and $J_{z\pm} = 0$ to $+1$ meV for significant trigonal distortion of $\Delta = +\lambda/2$. 

\subsection{Longer Range Couplings}
\begin{figure}[t]
\includegraphics[width=0.85\linewidth]{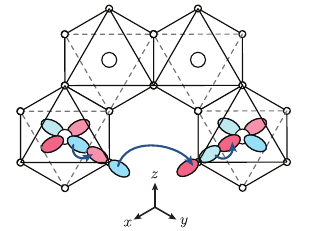}
\caption{3rd neighbor hopping relevant to $J_3$.}
\label{fig:hops3rd}
\end{figure}

While we have discussed above that $t_{2g}$-ligand hybridization should generally be small in $3d$ transition metal oxides (as reflected by small $t_{pd}^\pi$), the $e_g$-ligand hybridization may still play a significant role in the magnetic exchange due to the large $t_{pd}^\sigma$. This fact is what leads to significant ferromagnetic corrections $\delta J$ for nearest neighbor bonds. For longer range bonds, such as third neighbors in edge-sharing honeycomb lattices, it gives rise to a large hopping between $d_{x^2-y^2}$ orbitals shown in Fig.~\ref{fig:hops3rd} at order $(t_{pd}^\sigma)^2t_{pp}^\sigma/\Delta_{pd}^2 \sim 0.05$ to 0.1 eV. This is equivalent to a 3rd neighbor $t_5$, which allows the associated coupling to be readily estimated from the matrices $\mathbb{M}$. In particular, we estimate (for $\Delta_2 = 0$):
\begin{align}
J_3 \approx +0.5 \text{ to} +2.5\text{ meV}\\
K_3 \approx \Gamma_3 \approx 0
\end{align}
This is the only major third neighbor hopping pathway, so there are no additional terms to compete, and a relatively large antiferromagnetic $J_3$ should be expected for all honeycomb materials with partially occupied $e_g$ orbitals. For finite $\Delta_2$, the interactions are rendered anisotropic, with XXZ anisotropy reflecting the composition of the local moments. These estimates may be confirmed by considering the triangular lattice compound Ba$_3$CoSb$_2$O$_9$, for which the closest Co$^{2+}$ neighbors have the same geometry as depicted in Fig.~\ref{fig:hops3rd}. In this case, it is well established\cite{ma2016static,ito2017structure,ghioldi2018dynamical,kamiya2018nature} that the interactions are described by $J_{xy} = +1.7$ meV, $J_z = +1.5$ meV. Third neighbor interactions of similar magnitude should be expected for all honeycomb materials BCAO, BCPO, NCSO, and NCTO.

\section{Discussion}

\label{sec:materials}
In this work, we have considered the magnetic couplings in edge-sharing $d^7$ compounds. On this basis, we make several observations:

(1) All of the edge-sharing Co(II) oxides considered in this work appear to fall outside the regime of primary focus in previous theoretical studies\cite{liu2018pseudospin,doi:10.1142/S0217979221300061,liu2020kitaev,sano2018kitaev}. In particular, direct hopping likely dominates over ligand-assisted hopping ($t_3 \gg t_2$). In the realistic regime, we find that $K$ is generally suppressed compared to $J$, which calls into question models with dominant $K$ proposed for these materials. 

(2) The presence of the $e_g$ spins also opens additional exchange pathways. The $e_g$ orbitals mix relatively strongly with the ligand $p$-orbitals in contrast to the $t_{2g}$ orbitals. This enhances the importance of certain ligand exchange processes, which are responsible for ferromagnetic couplings in materials with 90$^\circ$ bond angles in the Goodenough-Kanamori description \cite{goodenough1963magnetism}. Analogous contributions are typically an order of magnitude smaller in low-spin $d^5$ compounds, such as A$_2$IrO$_3$ and $\alpha$-RuCl$_3$, where the $e_g$ orbitals are empty. The $e_g$ spins also participate in longer range antiferromagnetic couplings, corresponding to third neighbors for honeycomb lattice materials. It may be commonly expected that first and third neighbor interactions have similar magnitude in materials such as BCAO, BCPO, NCTO, and NCSO. Thus, while it was initially hoped that the smaller spatial extent of the $3d$ orbitals would suppress longer range coupling relative to those found in $4d^5$ and $5d^5$ compounds, the presence of $e_g$ spins likely enhances long-range interactions.

(3) Compared to heavy $d^5$ Kitaev materials such as iridates A$_2$IrO$_3$ and $\alpha$-RuCl$_3$, the weak spin-orbit coupling of Co increases the relative importance of local distortions. There are also many separate contributions to the magnetic couplings whose balance depends sensitively on local parameters such as $J_H, U$, and $\Delta_1$. This makes anticipating the magnetic Hamiltonian somewhat challenging. 
For Co(II) oxides, fortuitous fine-tuning may result in a different balance of couplings, but we anticipate that ferromagnetic $J$ (or equivalently $J_z, J_{xy}$) is likely always the largest coupling. The signs and magnitudes of the other couplings $K,\Gamma, \Gamma^\prime$ are influenced by the crystal field splitting and specific details of the hoppings. We find regions with all possible signs and relative magnitudes. 

(4) Real materials with small trigonal distortions (discussed in Section \ref{sec:geninteractions}) are likely described by $|K/J| \sim 0.2$ to 0.5, and $K \approx \Gamma$; specifically: $J \sim -8$ to $-2$ meV, $K \sim -2$ to $+2$ meV, and $\Gamma \sim -1$ to $+3$ meV. These magnitudes are compatible with the overall scale of interactions reported in the experimental literature\cite{regnault1977magnetic,nair2018short,regnault2018polarized,fava2020glide}. $\Gamma^\prime$ is predicted to be small unless there are significant departures from ideal symmetry of the bonds (i.e. $\Delta_2 = 0$, $C_{2h}$ bond symmetry, 90$^\circ$ M-L-M bond angles). It is not clear that a uniquely dominant $K$ is possible even for weak trigonal distortions.

(5) For systems with significant crystal field distortions (section \ref{sec:trig}), our findings are compatible with the historical description of Co(II) magnetic couplings in terms of XXZ models by M. E. Lines (Ref.~\onlinecite{lines1963magnetic}). This is true particularly because these considerations apply to the ferromagnetic ligand exchange processes, which we estimate are at least as important as processes involving $d$-$d$ hopping. In this case, the considerations discussed in Ref.~\onlinecite{liu2020kitaev} become equivalent to the classic results of Lines.
For trigonal crystal fields with $\Delta_2<0$ (corresponding to $g_{||}>g_{\perp}$), the Ising anisotropy induced by the crystal field may be very large ($J_z/J_{xy} \gg 1$), such that the couplings are dominated by a ferromagnetic $J_z$ with a common Ising axis for every bond. For positive crystal field $\Delta_2>0$ (corresponding to $g_{\perp}>g_{||}$), XXZ anisotropy is more limited ($J_{xy}/J_z \leq 4$), but may still be quite significant. Such materials are generally more desirable for realising strongly bond-dependent couplings, because the XXZ anisotropy is not as dominant.

(6) From the perspective of chemistry, it is unclear how to access the desirable ligand-assisted hopping regime where Kitaev coupling is largest. It is necessary to increase the metal-ligand hybridization relative to direct hopping between transition metal ions. This may typically be achieved by matching the electronegativity of the metals and ligands, such that $\Delta_{pd}$ is small. As a general trend, electronegativity of transition metals increases for heavier atoms, while the opposite is true for $p$-block ligand ions. The combination of heavy metals with heavy ligands results in the most covalent metal-ligand bonds. However, with increased covalency comes increased $t_{2g}$-$e_g$ splitting, and heavier metals tend to have reduced Coulomb terms $J_H$. As a result, the high-spin $(t_{2g})^5(e_g)^2$ state is only typically achievable in Co(II) compounds. By contrast, Rh(II) and Ni(III) tend to adopt low-spin $(t_{2g})^6(e_g)^1$ ground states. The most promising avenue then appears to be the combination of Co(II) with heavier ligands. With this in mind, we computed the hopping integrals for the triangular lattice compounds CoCl$_2$, CoBr$_2$ and CoI$_2$. For these compounds, we still estimate $|t_3 / t_2| \sim 4$. Thus, it is not clear that Kitaev-dominant exchange can be achieved without extreme fine tuning.

(7) Regarding NCSO and NCTO: Experimental estimates\cite{kim2021antiferromagnetic,liu2020kitaev,doi:10.1142/S0217979221300061} for NCSO and NCTO suggest $\Delta_2 \sim +4$ to $+13$ meV, which corresponds to a significant easy-plane XXZ anisotropy with $\delta \Gamma, \delta \Gamma^\prime \sim 0.2 - 0.5$ meV. Large departures from XXZ symmetry are also unlikely, given the small magnitude of the magnon gap ($\sim 1$ meV) observed in experiment at {\it both} the $\Gamma$-point\cite{lin2021field,chen2021spin} and the ordering wavevector\cite{songvilay2020kitaev,kim2021antiferromagnetic,chen2021spin,lin2021field}. It may further be remarked that the field-evolution of the ESR modes\cite{lin2021field} follow expectations for moderate easy-plane XXZ anisotropy. Thus, we may anticipate $J_1 < 0$, and $\Gamma,\Gamma^\prime > 0$, with the sign of $K$ being dependent on the microscopic details. 

It may be possible to place some constraints on the interactions on the basis of the ordered moment directions in the zigzag state. Related discussions appear in Ref.~\onlinecite{sanders2021dominant}. For NCTO, it was initially reported that the ordered moments lie nearly in the honeycomb plane, oriented along the direction of the ferromagnetic chains\cite{PhysRevB.94.214416}. This orientation was later disputed, with a significant tilting of the ordered moments out of the plane being reported in Ref.~\onlinecite{samarakoon2021static}. We may consider each case separately. 

If we consider the zigzag domain with magnetic wavevector parallel to the Z-bond, then moments oriented strictly in-plane would point along $\hat{e}_2^Z = (1,1,-2)/\sqrt{6}$ in cubic coordinates. 
The magnetic state would be left invariant under a 180$^\circ$ rotation around (111) followed by time reversal; it is thus reasonable to assume the couplings possess the same symmetry. This orientation generally requires\cite{chaloupka2016magnetic} $K,\Gamma >0$, and places the constraint on the couplings $\Gamma +\delta \Gamma= K+\Gamma^\prime + \delta \Gamma^\prime$ (i.e. $J_{z\pm} = 0$). Taken together, we suggest $J_1 = -3, J_3 = +2.5, K = \Gamma^\prime = +0.25, \Gamma = +0.5$ meV as an appropriate starting point for analysis. These considerations favor the model proposed in Ref.~\onlinecite{lin2021field}, and correspond to $J_{xy} = -3.25, J_z = -2.25, J_{\pm\pm} = -0.125, J_{z\pm} = 0$ meV. The antiferromagnetic sign of $K$ is possible with large $t_3$ and small $t_2>0$.

In contrast, if the moments are tilted out of plane, then fewer constraints may be placed on the couplings {\it a-priori}. However, such tilting likely requires $K < 0$. In this case, we suggest $J_1 = -3, J_3 = +2.5, K = -0.25, \Gamma^\prime = +0.25, \Gamma = +0.5$ meV as an appropriate starting point for analysis. The ferromagnetic sign of $K$ is possible with large $t_3$ and small $t_2<0$.

(8) Regarding BCAO: The breadth of experimental data on BCAO, in terms of the progression of field-induced phases\cite{huyan2022hydrostatic} and inelastic neutron data, provide a number of clues towards the magnetic model. While we leave full elaboration for future study\cite{maksimov2022}, some comments can be made. A recent reinvestigation\cite{regnault2018polarized} of the zero-field structure suggested it might better be described by a double stripe $\uparrow\uparrow \downarrow\downarrow$ analogue of the zigzag antiferromagnet, with moments oriented nearly along the in-plane $\hat{e}_2^Z$ direction, as with NCTO. This orientation points to $K,\Gamma>0$. The large discrepancy between in-plane critical fields (0.2, 0.5 T) and the out-of-plane critical  field (4T)\cite{zhang2021and} suggests significant anisotropy. Indeed, the $g$-tensor appears to satisfy\cite{regnault2018polarized} $g_{z} \sim 0.5 \ g_{xy}$, and within an XXZ model, $J_z \sim 0.4 \ J_{xy}$. These findings point to significant crystal field effects, with $\Delta_2 \sim 0.2$ to $0.25 \ \lambda$, i.e. $\Delta_2 \sim 15$ meV, implying significant $\Gamma$ and $\Gamma^\prime$ in the global coordinate scheme. In the XXZ scheme, the nearly in-plane moments suggest small $J_{z\pm}$, while an apparently small anisotropy between in-plane field directions\cite{regnault2018polarized} may place restrictions on $J_{\pm\pm}$. In contrast, the authors of Ref.~\onlinecite{zhang2021and} have advocated for small $J$, large $K<0$, and small average off-diagonal coupling $\bar{\Gamma}=(\Gamma +2\Gamma^\prime)/3$ on the basis of THz spectroscopy experiments. It should be emphasized that these conditions are not mutually compatible: small $J_{z\pm}$ and $J_{\pm\pm}$ implies small $K$, and large anisotropy between $J_{xy}$ and $J_z$ implies large $\Gamma+2\Gamma^\prime > 0$. If we consider BCAO to be in the physical regime of hoppings, our findings contradict the Kitaev-dominant model. We propose a model similar to NCTO with in-plane moments: $K$ is small and likely antiferromagnetic, $J <0$ is the dominant coupling, and $\Gamma, \Gamma^\prime > 0$ reflect a planar XY-anisotropy $|J_{xy}| > |J_z|$. The anomalous aspects of the ground state are then understood as a competition between $J_1, J_2$, and $J_3$, as previously suggested\cite{regnault1977magnetic,nair2018short,regnault2018polarized}. These suggestions are compatible with the recent {\it ab-initio} estimates\cite{das2021xy,maksimov2022}.

\section{Conclusions}
In high-spin Co(II) $3d^7$, the spin-orbital nature of the $j_{\rm eff} = 1/2$ ground state results in complex bond-dependent magnetic couplings in edge-sharing compounds. The theoretical description of these interactions is complicated by the existence of a wide number of different contributions, the balance of which is difficult to predict without detailed knowledge of the microscopic hopping, Coulomb, and crystal field parameters. Based on DFT studies of a range of Co oxides and halides, we find particular discrepancies with previous assumptions regarding these microscopic parameters. Specifically, direct metal-metal hopping is more prominent than ligand-assisted hopping, which results in significant suppression, and possible sign-reversal of the Kitaev interactions. Ferromagnetic exchange resulting from $e_g$ spins likely strongly dominates the nearest neighbor exchange. Off-diagonal couplings are likely enhanced by crystal field effects, resulting in significant XXZ anisotropy that is consistent with previous description of materials like the Ising ferromagnet CoNb$_2$O$_6$. Further neighbor exchange is also promoted by the presence of the $e_g$ spins. Taken together, Co(II) materials will continue to represent interesting platforms for highly complex anisotropic magnetism. However, the realisation of Kitaev magnetism specifically appears to be unrealistic from the microscopic perspective.

\section{Acknowledgements} 

We acknowledge useful discussions with D. Smirnov, Y. Jiang, S. Streltsov, P. Maksimov, R. Valenti, and E. Gati. We also thank P. Dai for bringing the problem to our attention. This work was supported by a pilot grant from the Center for Functional Materials. 
Computations were performed on the Wake Forest University DEAC Cluster, a centrally 
managed resource with support provided in part by Wake Forest University.

\bibliography{Co}

\begin{thebibliography}{81}%
\makeatletter
\providecommand \@ifxundefined [1]{%
 \@ifx{#1\undefined}
}%
\providecommand \@ifnum [1]{%
 \ifnum #1\expandafter \@firstoftwo
 \else \expandafter \@secondoftwo
 \fi
}%
\providecommand \@ifx [1]{%
 \ifx #1\expandafter \@firstoftwo
 \else \expandafter \@secondoftwo
 \fi
}%
\providecommand \natexlab [1]{#1}%
\providecommand \enquote  [1]{``#1''}%
\providecommand \bibnamefont  [1]{#1}%
\providecommand \bibfnamefont [1]{#1}%
\providecommand \citenamefont [1]{#1}%
\providecommand \href@noop [0]{\@secondoftwo}%
\providecommand \href [0]{\begingroup \@sanitize@url \@href}%
\providecommand \@href[1]{\@@startlink{#1}\@@href}%
\providecommand \@@href[1]{\endgroup#1\@@endlink}%
\providecommand \@sanitize@url [0]{\catcode `\\12\catcode `\$12\catcode
  `\&12\catcode `\#12\catcode `\^12\catcode `\_12\catcode `\%12\relax}%
\providecommand \@@startlink[1]{}%
\providecommand \@@endlink[0]{}%
\providecommand \url  [0]{\begingroup\@sanitize@url \@url }%
\providecommand \@url [1]{\endgroup\@href {#1}{\urlprefix }}%
\providecommand \urlprefix  [0]{URL }%
\providecommand \Eprint [0]{\href }%
\providecommand \doibase [0]{http://dx.doi.org/}%
\providecommand \selectlanguage [0]{\@gobble}%
\providecommand \bibinfo  [0]{\@secondoftwo}%
\providecommand \bibfield  [0]{\@secondoftwo}%
\providecommand \translation [1]{[#1]}%
\providecommand \BibitemOpen [0]{}%
\providecommand \bibitemStop [0]{}%
\providecommand \bibitemNoStop [0]{.\EOS\space}%
\providecommand \EOS [0]{\spacefactor3000\relax}%
\providecommand \BibitemShut  [1]{\csname bibitem#1\endcsname}%
\let\auto@bib@innerbib\@empty
%</preamble>
\bibitem [{\citenamefont {Nussinov}\ and\ \citenamefont {Van
  Den~Brink}(2015)}]{nussinov2015compass}%
  \BibitemOpen
  \bibfield  {author} {\bibinfo {author} {\bibfnamefont {Z.}~\bibnamefont
  {Nussinov}}\ and\ \bibinfo {author} {\bibfnamefont {J.}~\bibnamefont {Van
  Den~Brink}},\ }\href@noop {} {\bibfield  {journal} {\bibinfo  {journal} {Rev.
  Mod. Phys.}\ }\textbf {\bibinfo {volume} {87}},\ \bibinfo {pages} {1}
  (\bibinfo {year} {2015})}\BibitemShut {NoStop}%
\bibitem [{\citenamefont {Kitaev}(2006)}]{kitaev2006anyons}%
  \BibitemOpen
  \bibfield  {author} {\bibinfo {author} {\bibfnamefont {A.}~\bibnamefont
  {Kitaev}},\ }\href@noop {} {\bibfield  {journal} {\bibinfo  {journal} {Ann.
  Phys.}\ }\textbf {\bibinfo {volume} {321}},\ \bibinfo {pages} {2} (\bibinfo
  {year} {2006})}\BibitemShut {NoStop}%
\bibitem [{\citenamefont {Hermanns}\ \emph {et~al.}(2018)\citenamefont
  {Hermanns}, \citenamefont {Kimchi},\ and\ \citenamefont
  {Knolle}}]{hermanns2018physics}%
  \BibitemOpen
  \bibfield  {author} {\bibinfo {author} {\bibfnamefont {M.}~\bibnamefont
  {Hermanns}}, \bibinfo {author} {\bibfnamefont {I.}~\bibnamefont {Kimchi}}, \
  and\ \bibinfo {author} {\bibfnamefont {J.}~\bibnamefont {Knolle}},\
  }\href@noop {} {\bibfield  {journal} {\bibinfo  {journal} {Annu. Rev.
  Condens. Matter Phys.}\ }\textbf {\bibinfo {volume} {9}},\ \bibinfo {pages}
  {17} (\bibinfo {year} {2018})}\BibitemShut {NoStop}%
\bibitem [{\citenamefont {Broholm}\ \emph {et~al.}(2020)\citenamefont
  {Broholm}, \citenamefont {Cava}, \citenamefont {Kivelson}, \citenamefont
  {Nocera}, \citenamefont {Norman},\ and\ \citenamefont
  {Senthil}}]{broholm2020quantum}%
  \BibitemOpen
  \bibfield  {author} {\bibinfo {author} {\bibfnamefont {C.}~\bibnamefont
  {Broholm}}, \bibinfo {author} {\bibfnamefont {R.}~\bibnamefont {Cava}},
  \bibinfo {author} {\bibfnamefont {S.}~\bibnamefont {Kivelson}}, \bibinfo
  {author} {\bibfnamefont {D.}~\bibnamefont {Nocera}}, \bibinfo {author}
  {\bibfnamefont {M.}~\bibnamefont {Norman}}, \ and\ \bibinfo {author}
  {\bibfnamefont {T.}~\bibnamefont {Senthil}},\ }\href@noop {} {\bibfield
  {journal} {\bibinfo  {journal} {Science}\ }\textbf {\bibinfo {volume}
  {367}},\ \bibinfo {pages} {eaay0668} (\bibinfo {year} {2020})}\BibitemShut
  {NoStop}%
\bibitem [{\citenamefont {Zhou}\ \emph {et~al.}(2017)\citenamefont {Zhou},
  \citenamefont {Kanoda},\ and\ \citenamefont {Ng}}]{zhou2017quantum}%
  \BibitemOpen
  \bibfield  {author} {\bibinfo {author} {\bibfnamefont {Y.}~\bibnamefont
  {Zhou}}, \bibinfo {author} {\bibfnamefont {K.}~\bibnamefont {Kanoda}}, \ and\
  \bibinfo {author} {\bibfnamefont {T.-K.}\ \bibnamefont {Ng}},\ }\href@noop {}
  {\bibfield  {journal} {\bibinfo  {journal} {Rev. Mod. Phys.}\ }\textbf
  {\bibinfo {volume} {89}},\ \bibinfo {pages} {025003} (\bibinfo {year}
  {2017})}\BibitemShut {NoStop}%
\bibitem [{\citenamefont {Jackeli}\ and\ \citenamefont
  {Khaliullin}(2009)}]{jackeli2009mott}%
  \BibitemOpen
  \bibfield  {author} {\bibinfo {author} {\bibfnamefont {G.}~\bibnamefont
  {Jackeli}}\ and\ \bibinfo {author} {\bibfnamefont {G.}~\bibnamefont
  {Khaliullin}},\ }\href@noop {} {\bibfield  {journal} {\bibinfo  {journal}
  {Phys. Rev. Lett.}\ }\textbf {\bibinfo {volume} {102}},\ \bibinfo {pages}
  {017205} (\bibinfo {year} {2009})}\BibitemShut {NoStop}%
\bibitem [{\citenamefont {Winter}\ \emph
  {et~al.}(2017{\natexlab{a}})\citenamefont {Winter}, \citenamefont {Tsirlin},
  \citenamefont {Daghofer}, \citenamefont {van~den Brink}, \citenamefont
  {Singh}, \citenamefont {Gegenwart},\ and\ \citenamefont
  {Valenti}}]{winter2017models}%
  \BibitemOpen
  \bibfield  {author} {\bibinfo {author} {\bibfnamefont {S.~M.}\ \bibnamefont
  {Winter}}, \bibinfo {author} {\bibfnamefont {A.~A.}\ \bibnamefont {Tsirlin}},
  \bibinfo {author} {\bibfnamefont {M.}~\bibnamefont {Daghofer}}, \bibinfo
  {author} {\bibfnamefont {J.}~\bibnamefont {van~den Brink}}, \bibinfo {author}
  {\bibfnamefont {Y.}~\bibnamefont {Singh}}, \bibinfo {author} {\bibfnamefont
  {P.}~\bibnamefont {Gegenwart}}, \ and\ \bibinfo {author} {\bibfnamefont
  {R.}~\bibnamefont {Valenti}},\ }\href@noop {} {\bibfield  {journal} {\bibinfo
   {journal} {J. Phys. Condens. Matter}\ }\textbf {\bibinfo {volume} {29}},\
  \bibinfo {pages} {493002} (\bibinfo {year} {2017}{\natexlab{a}})}\BibitemShut
  {NoStop}%
\bibitem [{\citenamefont {Trebst}(2017)}]{trebst2017kitaev}%
  \BibitemOpen
  \bibfield  {author} {\bibinfo {author} {\bibfnamefont {S.}~\bibnamefont
  {Trebst}},\ }\href@noop {} {\bibfield  {journal} {\bibinfo  {journal} {arXiv
  preprint arXiv:1701.07056}\ } (\bibinfo {year} {2017})}\BibitemShut {NoStop}%
\bibitem [{\citenamefont {Singh}\ and\ \citenamefont
  {Gegenwart}(2010)}]{singh2010antiferromagnetic}%
  \BibitemOpen
  \bibfield  {author} {\bibinfo {author} {\bibfnamefont {Y.}~\bibnamefont
  {Singh}}\ and\ \bibinfo {author} {\bibfnamefont {P.}~\bibnamefont
  {Gegenwart}},\ }\href@noop {} {\bibfield  {journal} {\bibinfo  {journal}
  {Phys. Rev. B}\ }\textbf {\bibinfo {volume} {82}},\ \bibinfo {pages} {064412}
  (\bibinfo {year} {2010})}\BibitemShut {NoStop}%
\bibitem [{\citenamefont {Plumb}\ \emph {et~al.}(2014)\citenamefont {Plumb},
  \citenamefont {Clancy}, \citenamefont {Sandilands}, \citenamefont {Shankar},
  \citenamefont {Hu}, \citenamefont {Burch}, \citenamefont {Kee},\ and\
  \citenamefont {Kim}}]{plumb2014alpha}%
  \BibitemOpen
  \bibfield  {author} {\bibinfo {author} {\bibfnamefont {K.}~\bibnamefont
  {Plumb}}, \bibinfo {author} {\bibfnamefont {J.}~\bibnamefont {Clancy}},
  \bibinfo {author} {\bibfnamefont {L.}~\bibnamefont {Sandilands}}, \bibinfo
  {author} {\bibfnamefont {V.~V.}\ \bibnamefont {Shankar}}, \bibinfo {author}
  {\bibfnamefont {Y.}~\bibnamefont {Hu}}, \bibinfo {author} {\bibfnamefont
  {K.}~\bibnamefont {Burch}}, \bibinfo {author} {\bibfnamefont {H.-Y.}\
  \bibnamefont {Kee}}, \ and\ \bibinfo {author} {\bibfnamefont {Y.-J.}\
  \bibnamefont {Kim}},\ }\href@noop {} {\bibfield  {journal} {\bibinfo
  {journal} {Phys. Rev. B}\ }\textbf {\bibinfo {volume} {90}},\ \bibinfo
  {pages} {041112} (\bibinfo {year} {2014})}\BibitemShut {NoStop}%
\bibitem [{\citenamefont {Banerjee}\ \emph {et~al.}(2016)\citenamefont
  {Banerjee}, \citenamefont {Bridges}, \citenamefont {Yan}, \citenamefont
  {Aczel}, \citenamefont {Li}, \citenamefont {Stone}, \citenamefont {Granroth},
  \citenamefont {Lumsden}, \citenamefont {Yiu}, \citenamefont {Knolle},
  \citenamefont {Bhattacharjee}, \citenamefont {Kovrizhin}, \citenamefont
  {Moessner}, \citenamefont {Tennant}, \citenamefont {Mandrus},\ and\
  \citenamefont {Nagler}}]{banerjee2016proximate}%
  \BibitemOpen
  \bibfield  {author} {\bibinfo {author} {\bibfnamefont {A.}~\bibnamefont
  {Banerjee}}, \bibinfo {author} {\bibfnamefont {C.}~\bibnamefont {Bridges}},
  \bibinfo {author} {\bibfnamefont {J.-Q.}\ \bibnamefont {Yan}}, \bibinfo
  {author} {\bibfnamefont {A.}~\bibnamefont {Aczel}}, \bibinfo {author}
  {\bibfnamefont {L.}~\bibnamefont {Li}}, \bibinfo {author} {\bibfnamefont
  {M.}~\bibnamefont {Stone}}, \bibinfo {author} {\bibfnamefont
  {G.}~\bibnamefont {Granroth}}, \bibinfo {author} {\bibfnamefont
  {M.}~\bibnamefont {Lumsden}}, \bibinfo {author} {\bibfnamefont
  {Y.}~\bibnamefont {Yiu}}, \bibinfo {author} {\bibfnamefont {J.}~\bibnamefont
  {Knolle}}, \bibinfo {author} {\bibfnamefont {S.}~\bibnamefont
  {Bhattacharjee}}, \bibinfo {author} {\bibfnamefont {D.~L.}\ \bibnamefont
  {Kovrizhin}}, \bibinfo {author} {\bibfnamefont {R.}~\bibnamefont {Moessner}},
  \bibinfo {author} {\bibfnamefont {D.~A.}\ \bibnamefont {Tennant}}, \bibinfo
  {author} {\bibfnamefont {D.~G.}\ \bibnamefont {Mandrus}}, \ and\ \bibinfo
  {author} {\bibfnamefont {S.~E.}\ \bibnamefont {Nagler}},\ }\href@noop {}
  {\bibfield  {journal} {\bibinfo  {journal} {Nat. Mater.}\ }\textbf {\bibinfo
  {volume} {15}},\ \bibinfo {pages} {733} (\bibinfo {year} {2016})}\BibitemShut
  {NoStop}%
\bibitem [{\citenamefont {Banerjee}\ \emph {et~al.}(2017)\citenamefont
  {Banerjee}, \citenamefont {Yan}, \citenamefont {Knolle}, \citenamefont
  {Bridges}, \citenamefont {Stone}, \citenamefont {Lumsden}, \citenamefont
  {Mandrus}, \citenamefont {Tennant}, \citenamefont {Moessner},\ and\
  \citenamefont {Nagler}}]{banerjee2017neutron}%
  \BibitemOpen
  \bibfield  {author} {\bibinfo {author} {\bibfnamefont {A.}~\bibnamefont
  {Banerjee}}, \bibinfo {author} {\bibfnamefont {J.}~\bibnamefont {Yan}},
  \bibinfo {author} {\bibfnamefont {J.}~\bibnamefont {Knolle}}, \bibinfo
  {author} {\bibfnamefont {C.~A.}\ \bibnamefont {Bridges}}, \bibinfo {author}
  {\bibfnamefont {M.~B.}\ \bibnamefont {Stone}}, \bibinfo {author}
  {\bibfnamefont {M.~D.}\ \bibnamefont {Lumsden}}, \bibinfo {author}
  {\bibfnamefont {D.~G.}\ \bibnamefont {Mandrus}}, \bibinfo {author}
  {\bibfnamefont {D.~A.}\ \bibnamefont {Tennant}}, \bibinfo {author}
  {\bibfnamefont {R.}~\bibnamefont {Moessner}}, \ and\ \bibinfo {author}
  {\bibfnamefont {S.~E.}\ \bibnamefont {Nagler}},\ }\href@noop {} {\bibfield
  {journal} {\bibinfo  {journal} {Science}\ }\textbf {\bibinfo {volume}
  {356}},\ \bibinfo {pages} {1055} (\bibinfo {year} {2017})}\BibitemShut
  {NoStop}%
\bibitem [{\citenamefont {Hwan~Chun}\ \emph {et~al.}(2015)\citenamefont
  {Hwan~Chun}, \citenamefont {Kim}, \citenamefont {Kim}, \citenamefont {Zheng},
  \citenamefont {Stoumpos}, \citenamefont {Malliakas}, \citenamefont
  {Mitchell}, \citenamefont {Mehlawat}, \citenamefont {Singh}, \citenamefont
  {Choi}, \citenamefont {Gog}, \citenamefont {Al-Zein}, \citenamefont {Sala},
  \citenamefont {Krisch}, \citenamefont {Chaloupka}, \citenamefont {Jackeli},
  \citenamefont {Khaliullin},\ and\ \citenamefont {Kim}}]{hwan2015direct}%
  \BibitemOpen
  \bibfield  {author} {\bibinfo {author} {\bibfnamefont {S.}~\bibnamefont
  {Hwan~Chun}}, \bibinfo {author} {\bibfnamefont {J.-W.}\ \bibnamefont {Kim}},
  \bibinfo {author} {\bibfnamefont {J.}~\bibnamefont {Kim}}, \bibinfo {author}
  {\bibfnamefont {H.}~\bibnamefont {Zheng}}, \bibinfo {author} {\bibfnamefont
  {C.~C.}\ \bibnamefont {Stoumpos}}, \bibinfo {author} {\bibfnamefont
  {C.}~\bibnamefont {Malliakas}}, \bibinfo {author} {\bibfnamefont
  {J.}~\bibnamefont {Mitchell}}, \bibinfo {author} {\bibfnamefont
  {K.}~\bibnamefont {Mehlawat}}, \bibinfo {author} {\bibfnamefont
  {Y.}~\bibnamefont {Singh}}, \bibinfo {author} {\bibfnamefont
  {Y.}~\bibnamefont {Choi}}, \bibinfo {author} {\bibfnamefont {T.}~\bibnamefont
  {Gog}}, \bibinfo {author} {\bibfnamefont {A.}~\bibnamefont {Al-Zein}},
  \bibinfo {author} {\bibfnamefont {M.~M.}\ \bibnamefont {Sala}}, \bibinfo
  {author} {\bibfnamefont {M.}~\bibnamefont {Krisch}}, \bibinfo {author}
  {\bibfnamefont {J.}~\bibnamefont {Chaloupka}}, \bibinfo {author}
  {\bibfnamefont {G.}~\bibnamefont {Jackeli}}, \bibinfo {author} {\bibfnamefont
  {G.}~\bibnamefont {Khaliullin}}, \ and\ \bibinfo {author} {\bibfnamefont
  {B.~J.}\ \bibnamefont {Kim}},\ }\href@noop {} {\bibfield  {journal} {\bibinfo
   {journal} {Nat. Phys.}\ }\textbf {\bibinfo {volume} {11}},\ \bibinfo {pages}
  {462} (\bibinfo {year} {2015})}\BibitemShut {NoStop}%
\bibitem [{\citenamefont {Suzuki}\ \emph {et~al.}(2021)\citenamefont {Suzuki},
  \citenamefont {Liu}, \citenamefont {Bertinshaw}, \citenamefont {Ueda},
  \citenamefont {Kim}, \citenamefont {Laha}, \citenamefont {Weber},
  \citenamefont {Yang}, \citenamefont {Wang}, \citenamefont {Takahashi},
  \citenamefont {Fürsich}, \citenamefont {Minola}, \citenamefont {Lotsch},
  \citenamefont {Kim}, \citenamefont {Yavas}, \citenamefont {Daghofer},
  \citenamefont {Chaloupka}, \citenamefont {Khaliullin}, \citenamefont
  {Gretarsson},\ and\ \citenamefont {Keimer}}]{suzuki2021proximate}%
  \BibitemOpen
  \bibfield  {author} {\bibinfo {author} {\bibfnamefont {H.}~\bibnamefont
  {Suzuki}}, \bibinfo {author} {\bibfnamefont {H.}~\bibnamefont {Liu}},
  \bibinfo {author} {\bibfnamefont {J.}~\bibnamefont {Bertinshaw}}, \bibinfo
  {author} {\bibfnamefont {K.}~\bibnamefont {Ueda}}, \bibinfo {author}
  {\bibfnamefont {H.}~\bibnamefont {Kim}}, \bibinfo {author} {\bibfnamefont
  {S.}~\bibnamefont {Laha}}, \bibinfo {author} {\bibfnamefont {D.}~\bibnamefont
  {Weber}}, \bibinfo {author} {\bibfnamefont {Z.}~\bibnamefont {Yang}},
  \bibinfo {author} {\bibfnamefont {L.}~\bibnamefont {Wang}}, \bibinfo {author}
  {\bibfnamefont {H.}~\bibnamefont {Takahashi}}, \bibinfo {author}
  {\bibfnamefont {K.}~\bibnamefont {Fürsich}}, \bibinfo {author}
  {\bibfnamefont {M.}~\bibnamefont {Minola}}, \bibinfo {author} {\bibfnamefont
  {B.~V.}\ \bibnamefont {Lotsch}}, \bibinfo {author} {\bibfnamefont {B.~J.}\
  \bibnamefont {Kim}}, \bibinfo {author} {\bibfnamefont {H.}~\bibnamefont
  {Yavas}}, \bibinfo {author} {\bibfnamefont {M.}~\bibnamefont {Daghofer}},
  \bibinfo {author} {\bibfnamefont {J.}~\bibnamefont {Chaloupka}}, \bibinfo
  {author} {\bibfnamefont {G.}~\bibnamefont {Khaliullin}}, \bibinfo {author}
  {\bibfnamefont {H.}~\bibnamefont {Gretarsson}}, \ and\ \bibinfo {author}
  {\bibfnamefont {B.}~\bibnamefont {Keimer}},\ }\href@noop {} {\bibfield
  {journal} {\bibinfo  {journal} {Nat. Commun.}\ }\textbf {\bibinfo {volume}
  {12}},\ \bibinfo {pages} {1} (\bibinfo {year} {2021})}\BibitemShut {NoStop}%
\bibitem [{\citenamefont {Winter}\ \emph
  {et~al.}(2017{\natexlab{b}})\citenamefont {Winter}, \citenamefont {Riedl},
  \citenamefont {Maksimov}, \citenamefont {Chernyshev}, \citenamefont
  {Honecker},\ and\ \citenamefont {Valent{\'\i}}}]{winter2017breakdown}%
  \BibitemOpen
  \bibfield  {author} {\bibinfo {author} {\bibfnamefont {S.~M.}\ \bibnamefont
  {Winter}}, \bibinfo {author} {\bibfnamefont {K.}~\bibnamefont {Riedl}},
  \bibinfo {author} {\bibfnamefont {P.~A.}\ \bibnamefont {Maksimov}}, \bibinfo
  {author} {\bibfnamefont {A.~L.}\ \bibnamefont {Chernyshev}}, \bibinfo
  {author} {\bibfnamefont {A.}~\bibnamefont {Honecker}}, \ and\ \bibinfo
  {author} {\bibfnamefont {R.}~\bibnamefont {Valent{\'\i}}},\ }\href@noop {}
  {\bibfield  {journal} {\bibinfo  {journal} {Nat. Commun.}\ }\textbf {\bibinfo
  {volume} {8}},\ \bibinfo {pages} {1} (\bibinfo {year}
  {2017}{\natexlab{b}})}\BibitemShut {NoStop}%
\bibitem [{\citenamefont {Banerjee}\ \emph {et~al.}(2018)\citenamefont
  {Banerjee}, \citenamefont {Lampen-Kelley}, \citenamefont {Knolle},
  \citenamefont {Balz}, \citenamefont {Aczel}, \citenamefont {Winn},
  \citenamefont {Liu}, \citenamefont {Pajerowski}, \citenamefont {Yan},
  \citenamefont {Bridges}, \citenamefont {Savici}, \citenamefont {Chakoumakos},
  \citenamefont {Lumsden}, \citenamefont {Tennant}, \citenamefont {Moessner},
  \citenamefont {Mandrus},\ and\ \citenamefont
  {Nagler}}]{banerjee2018excitations}%
  \BibitemOpen
  \bibfield  {author} {\bibinfo {author} {\bibfnamefont {A.}~\bibnamefont
  {Banerjee}}, \bibinfo {author} {\bibfnamefont {P.}~\bibnamefont
  {Lampen-Kelley}}, \bibinfo {author} {\bibfnamefont {J.}~\bibnamefont
  {Knolle}}, \bibinfo {author} {\bibfnamefont {C.}~\bibnamefont {Balz}},
  \bibinfo {author} {\bibfnamefont {A.~A.}\ \bibnamefont {Aczel}}, \bibinfo
  {author} {\bibfnamefont {B.}~\bibnamefont {Winn}}, \bibinfo {author}
  {\bibfnamefont {Y.}~\bibnamefont {Liu}}, \bibinfo {author} {\bibfnamefont
  {D.}~\bibnamefont {Pajerowski}}, \bibinfo {author} {\bibfnamefont
  {J.}~\bibnamefont {Yan}}, \bibinfo {author} {\bibfnamefont {C.~A.}\
  \bibnamefont {Bridges}}, \bibinfo {author} {\bibfnamefont {A.~T.}\
  \bibnamefont {Savici}}, \bibinfo {author} {\bibfnamefont {B.~C.}\
  \bibnamefont {Chakoumakos}}, \bibinfo {author} {\bibfnamefont {M.~D.}\
  \bibnamefont {Lumsden}}, \bibinfo {author} {\bibfnamefont {D.~A.}\
  \bibnamefont {Tennant}}, \bibinfo {author} {\bibfnamefont {R.}~\bibnamefont
  {Moessner}}, \bibinfo {author} {\bibfnamefont {D.~G.}\ \bibnamefont
  {Mandrus}}, \ and\ \bibinfo {author} {\bibfnamefont {S.~E.}\ \bibnamefont
  {Nagler}},\ }\href@noop {} {\bibfield  {journal} {\bibinfo  {journal} {npj
  Quantum Mater.}\ }\textbf {\bibinfo {volume} {3}},\ \bibinfo {pages} {1}
  (\bibinfo {year} {2018})}\BibitemShut {NoStop}%
\bibitem [{\citenamefont {Kasahara}\ \emph {et~al.}(2018)\citenamefont
  {Kasahara}, \citenamefont {Ohnishi}, \citenamefont {Mizukami}, \citenamefont
  {Tanaka}, \citenamefont {Ma}, \citenamefont {Sugii}, \citenamefont {Kurita},
  \citenamefont {Tanaka}, \citenamefont {Nasu}, \citenamefont {Motome},
  \citenamefont {Shibauchi},\ and\ \citenamefont
  {Matsuda}}]{kasahara2018majorana}%
  \BibitemOpen
  \bibfield  {author} {\bibinfo {author} {\bibfnamefont {Y.}~\bibnamefont
  {Kasahara}}, \bibinfo {author} {\bibfnamefont {T.}~\bibnamefont {Ohnishi}},
  \bibinfo {author} {\bibfnamefont {Y.}~\bibnamefont {Mizukami}}, \bibinfo
  {author} {\bibfnamefont {O.}~\bibnamefont {Tanaka}}, \bibinfo {author}
  {\bibfnamefont {S.}~\bibnamefont {Ma}}, \bibinfo {author} {\bibfnamefont
  {K.}~\bibnamefont {Sugii}}, \bibinfo {author} {\bibfnamefont
  {N.}~\bibnamefont {Kurita}}, \bibinfo {author} {\bibfnamefont
  {H.}~\bibnamefont {Tanaka}}, \bibinfo {author} {\bibfnamefont
  {J.}~\bibnamefont {Nasu}}, \bibinfo {author} {\bibfnamefont {Y.}~\bibnamefont
  {Motome}}, \bibinfo {author} {\bibfnamefont {T.}~\bibnamefont {Shibauchi}}, \
  and\ \bibinfo {author} {\bibfnamefont {Y.}~\bibnamefont {Matsuda}},\
  }\href@noop {} {\bibfield  {journal} {\bibinfo  {journal} {Nature}\ }\textbf
  {\bibinfo {volume} {559}},\ \bibinfo {pages} {227} (\bibinfo {year}
  {2018})}\BibitemShut {NoStop}%
\bibitem [{\citenamefont {Yokoi}\ \emph {et~al.}(2021)\citenamefont {Yokoi},
  \citenamefont {Ma}, \citenamefont {Kasahara}, \citenamefont {Kasahara},
  \citenamefont {Shibauchi}, \citenamefont {Kurita}, \citenamefont {Tanaka},
  \citenamefont {Nasu}, \citenamefont {Motome}, \citenamefont {Hickey},
  \citenamefont {Trebst},\ and\ \citenamefont {Matsuda}}]{yokoi2021half}%
  \BibitemOpen
  \bibfield  {author} {\bibinfo {author} {\bibfnamefont {T.}~\bibnamefont
  {Yokoi}}, \bibinfo {author} {\bibfnamefont {S.}~\bibnamefont {Ma}}, \bibinfo
  {author} {\bibfnamefont {Y.}~\bibnamefont {Kasahara}}, \bibinfo {author}
  {\bibfnamefont {S.}~\bibnamefont {Kasahara}}, \bibinfo {author}
  {\bibfnamefont {T.}~\bibnamefont {Shibauchi}}, \bibinfo {author}
  {\bibfnamefont {N.}~\bibnamefont {Kurita}}, \bibinfo {author} {\bibfnamefont
  {H.}~\bibnamefont {Tanaka}}, \bibinfo {author} {\bibfnamefont
  {J.}~\bibnamefont {Nasu}}, \bibinfo {author} {\bibfnamefont {Y.}~\bibnamefont
  {Motome}}, \bibinfo {author} {\bibfnamefont {C.}~\bibnamefont {Hickey}},
  \bibinfo {author} {\bibfnamefont {S.}~\bibnamefont {Trebst}}, \ and\ \bibinfo
  {author} {\bibfnamefont {Y.}~\bibnamefont {Matsuda}},\ }\href@noop {}
  {\bibfield  {journal} {\bibinfo  {journal} {Science}\ }\textbf {\bibinfo
  {volume} {373}},\ \bibinfo {pages} {568} (\bibinfo {year}
  {2021})}\BibitemShut {NoStop}%
\bibitem [{\citenamefont {Liu}\ and\ \citenamefont
  {Khaliullin}(2018)}]{liu2018pseudospin}%
  \BibitemOpen
  \bibfield  {author} {\bibinfo {author} {\bibfnamefont {H.}~\bibnamefont
  {Liu}}\ and\ \bibinfo {author} {\bibfnamefont {G.}~\bibnamefont
  {Khaliullin}},\ }\href@noop {} {\bibfield  {journal} {\bibinfo  {journal}
  {Phys. Rev. B}\ }\textbf {\bibinfo {volume} {97}},\ \bibinfo {pages} {014407}
  (\bibinfo {year} {2018})}\BibitemShut {NoStop}%
\bibitem [{\citenamefont {Liu}(2021)}]{doi:10.1142/S0217979221300061}%
  \BibitemOpen
  \bibfield  {author} {\bibinfo {author} {\bibfnamefont {H.}~\bibnamefont
  {Liu}},\ }\href {\doibase 10.1142/S0217979221300061} {\bibfield  {journal}
  {\bibinfo  {journal} {Int. J. Mod. Phys. B}\ }\textbf {\bibinfo {volume}
  {35}},\ \bibinfo {pages} {2130006} (\bibinfo {year} {2021})}\BibitemShut
  {NoStop}%
\bibitem [{\citenamefont {Liu}\ \emph {et~al.}(2020)\citenamefont {Liu},
  \citenamefont {Chaloupka},\ and\ \citenamefont {Khaliullin}}]{liu2020kitaev}%
  \BibitemOpen
  \bibfield  {author} {\bibinfo {author} {\bibfnamefont {H.}~\bibnamefont
  {Liu}}, \bibinfo {author} {\bibfnamefont {J.}~\bibnamefont {Chaloupka}}, \
  and\ \bibinfo {author} {\bibfnamefont {G.}~\bibnamefont {Khaliullin}},\
  }\href@noop {} {\bibfield  {journal} {\bibinfo  {journal} {Phys. Rev. Lett.}\
  }\textbf {\bibinfo {volume} {125}},\ \bibinfo {pages} {047201} (\bibinfo
  {year} {2020})}\BibitemShut {NoStop}%
\bibitem [{\citenamefont {Sano}\ \emph {et~al.}(2018)\citenamefont {Sano},
  \citenamefont {Kato},\ and\ \citenamefont {Motome}}]{sano2018kitaev}%
  \BibitemOpen
  \bibfield  {author} {\bibinfo {author} {\bibfnamefont {R.}~\bibnamefont
  {Sano}}, \bibinfo {author} {\bibfnamefont {Y.}~\bibnamefont {Kato}}, \ and\
  \bibinfo {author} {\bibfnamefont {Y.}~\bibnamefont {Motome}},\ }\href@noop {}
  {\bibfield  {journal} {\bibinfo  {journal} {Phys. Rev. B}\ }\textbf {\bibinfo
  {volume} {97}},\ \bibinfo {pages} {014408} (\bibinfo {year}
  {2018})}\BibitemShut {NoStop}%
\bibitem [{\citenamefont {Rau}\ \emph {et~al.}(2014)\citenamefont {Rau},
  \citenamefont {Lee},\ and\ \citenamefont {Kee}}]{rau2014generic}%
  \BibitemOpen
  \bibfield  {author} {\bibinfo {author} {\bibfnamefont {J.~G.}\ \bibnamefont
  {Rau}}, \bibinfo {author} {\bibfnamefont {E.~K.-H.}\ \bibnamefont {Lee}}, \
  and\ \bibinfo {author} {\bibfnamefont {H.-Y.}\ \bibnamefont {Kee}},\
  }\href@noop {} {\bibfield  {journal} {\bibinfo  {journal} {Phys. Rev. Lett.}\
  }\textbf {\bibinfo {volume} {112}},\ \bibinfo {pages} {077204} (\bibinfo
  {year} {2014})}\BibitemShut {NoStop}%
\bibitem [{\citenamefont {Winter}\ \emph {et~al.}(2016)\citenamefont {Winter},
  \citenamefont {Li}, \citenamefont {Jeschke},\ and\ \citenamefont
  {Valent{\'\i}}}]{winter2016challenges}%
  \BibitemOpen
  \bibfield  {author} {\bibinfo {author} {\bibfnamefont {S.~M.}\ \bibnamefont
  {Winter}}, \bibinfo {author} {\bibfnamefont {Y.}~\bibnamefont {Li}}, \bibinfo
  {author} {\bibfnamefont {H.~O.}\ \bibnamefont {Jeschke}}, \ and\ \bibinfo
  {author} {\bibfnamefont {R.}~\bibnamefont {Valent{\'\i}}},\ }\href@noop {}
  {\bibfield  {journal} {\bibinfo  {journal} {Phys. Rev. B}\ }\textbf {\bibinfo
  {volume} {93}},\ \bibinfo {pages} {214431} (\bibinfo {year}
  {2016})}\BibitemShut {NoStop}%
\bibitem [{\citenamefont {Lines}(1963)}]{lines1963magnetic}%
  \BibitemOpen
  \bibfield  {author} {\bibinfo {author} {\bibfnamefont {M.}~\bibnamefont
  {Lines}},\ }\href@noop {} {\bibfield  {journal} {\bibinfo  {journal} {Phys.
  Rev.}\ }\textbf {\bibinfo {volume} {131}},\ \bibinfo {pages} {546} (\bibinfo
  {year} {1963})}\BibitemShut {NoStop}%
\bibitem [{\citenamefont {Oguchi}(1965)}]{oguchi1965theory}%
  \BibitemOpen
  \bibfield  {author} {\bibinfo {author} {\bibfnamefont {T.}~\bibnamefont
  {Oguchi}},\ }\href@noop {} {\bibfield  {journal} {\bibinfo  {journal} {J.
  Phys. Soc. Japan}\ }\textbf {\bibinfo {volume} {20}},\ \bibinfo {pages}
  {2236} (\bibinfo {year} {1965})}\BibitemShut {NoStop}%
\bibitem [{\citenamefont {Scharf}\ \emph {et~al.}(1979)\citenamefont {Scharf},
  \citenamefont {Weitzel}, \citenamefont {Yaeger}, \citenamefont {Maartense},\
  and\ \citenamefont {Wanklyn}}]{scharf1979magnetic}%
  \BibitemOpen
  \bibfield  {author} {\bibinfo {author} {\bibfnamefont {W.}~\bibnamefont
  {Scharf}}, \bibinfo {author} {\bibfnamefont {H.}~\bibnamefont {Weitzel}},
  \bibinfo {author} {\bibfnamefont {I.}~\bibnamefont {Yaeger}}, \bibinfo
  {author} {\bibfnamefont {I.}~\bibnamefont {Maartense}}, \ and\ \bibinfo
  {author} {\bibfnamefont {B.}~\bibnamefont {Wanklyn}},\ }\href@noop {}
  {\bibfield  {journal} {\bibinfo  {journal} {J. Magn. Magn. Mater.}\ }\textbf
  {\bibinfo {volume} {13}},\ \bibinfo {pages} {121} (\bibinfo {year}
  {1979})}\BibitemShut {NoStop}%
\bibitem [{\citenamefont {Maartense}\ \emph {et~al.}(1977)\citenamefont
  {Maartense}, \citenamefont {Yaeger},\ and\ \citenamefont
  {Wanklyn}}]{maartense1977field}%
  \BibitemOpen
  \bibfield  {author} {\bibinfo {author} {\bibfnamefont {I.}~\bibnamefont
  {Maartense}}, \bibinfo {author} {\bibfnamefont {I.}~\bibnamefont {Yaeger}}, \
  and\ \bibinfo {author} {\bibfnamefont {B.}~\bibnamefont {Wanklyn}},\
  }\href@noop {} {\bibfield  {journal} {\bibinfo  {journal} {Solid State
  Commun.}\ }\textbf {\bibinfo {volume} {21}},\ \bibinfo {pages} {93} (\bibinfo
  {year} {1977})}\BibitemShut {NoStop}%
\bibitem [{\citenamefont {Kobayashi}\ \emph {et~al.}(1999)\citenamefont
  {Kobayashi}, \citenamefont {Mitsuda}, \citenamefont {Ishikawa}, \citenamefont
  {Miyatani},\ and\ \citenamefont {Kohn}}]{kobayashi1999three}%
  \BibitemOpen
  \bibfield  {author} {\bibinfo {author} {\bibfnamefont {S.}~\bibnamefont
  {Kobayashi}}, \bibinfo {author} {\bibfnamefont {S.}~\bibnamefont {Mitsuda}},
  \bibinfo {author} {\bibfnamefont {M.}~\bibnamefont {Ishikawa}}, \bibinfo
  {author} {\bibfnamefont {K.}~\bibnamefont {Miyatani}}, \ and\ \bibinfo
  {author} {\bibfnamefont {K.}~\bibnamefont {Kohn}},\ }\href@noop {} {\bibfield
   {journal} {\bibinfo  {journal} {Phys. Rev. B}\ }\textbf {\bibinfo {volume}
  {60}},\ \bibinfo {pages} {3331} (\bibinfo {year} {1999})}\BibitemShut
  {NoStop}%
\bibitem [{\citenamefont {Lee}\ \emph {et~al.}(2010)\citenamefont {Lee},
  \citenamefont {Kaul},\ and\ \citenamefont {Balents}}]{lee2010interplay}%
  \BibitemOpen
  \bibfield  {author} {\bibinfo {author} {\bibfnamefont {S.}~\bibnamefont
  {Lee}}, \bibinfo {author} {\bibfnamefont {R.~K.}\ \bibnamefont {Kaul}}, \
  and\ \bibinfo {author} {\bibfnamefont {L.}~\bibnamefont {Balents}},\
  }\href@noop {} {\bibfield  {journal} {\bibinfo  {journal} {Nat. Phys.}\
  }\textbf {\bibinfo {volume} {6}},\ \bibinfo {pages} {702} (\bibinfo {year}
  {2010})}\BibitemShut {NoStop}%
\bibitem [{\citenamefont {Coldea}\ \emph {et~al.}(2010)\citenamefont {Coldea},
  \citenamefont {Tennant}, \citenamefont {Wheeler}, \citenamefont {Wawrzynska},
  \citenamefont {Prabhakaran}, \citenamefont {Telling}, \citenamefont
  {Habicht}, \citenamefont {Smeibidl},\ and\ \citenamefont
  {Kiefer}}]{coldea2010quantum}%
  \BibitemOpen
  \bibfield  {author} {\bibinfo {author} {\bibfnamefont {R.}~\bibnamefont
  {Coldea}}, \bibinfo {author} {\bibfnamefont {D.}~\bibnamefont {Tennant}},
  \bibinfo {author} {\bibfnamefont {E.}~\bibnamefont {Wheeler}}, \bibinfo
  {author} {\bibfnamefont {E.}~\bibnamefont {Wawrzynska}}, \bibinfo {author}
  {\bibfnamefont {D.}~\bibnamefont {Prabhakaran}}, \bibinfo {author}
  {\bibfnamefont {M.}~\bibnamefont {Telling}}, \bibinfo {author} {\bibfnamefont
  {K.}~\bibnamefont {Habicht}}, \bibinfo {author} {\bibfnamefont
  {P.}~\bibnamefont {Smeibidl}}, \ and\ \bibinfo {author} {\bibfnamefont
  {K.}~\bibnamefont {Kiefer}},\ }\href@noop {} {\bibfield  {journal} {\bibinfo
  {journal} {Science}\ }\textbf {\bibinfo {volume} {327}},\ \bibinfo {pages}
  {177} (\bibinfo {year} {2010})}\BibitemShut {NoStop}%
\bibitem [{\citenamefont {Morris}\ \emph {et~al.}(2014)\citenamefont {Morris},
  \citenamefont {Aguilar}, \citenamefont {Ghosh}, \citenamefont {Koohpayeh},
  \citenamefont {Krizan}, \citenamefont {Cava}, \citenamefont {Tchernyshyov},
  \citenamefont {McQueen},\ and\ \citenamefont
  {Armitage}}]{morris2014hierarchy}%
  \BibitemOpen
  \bibfield  {author} {\bibinfo {author} {\bibfnamefont {C.}~\bibnamefont
  {Morris}}, \bibinfo {author} {\bibfnamefont {R.~V.}\ \bibnamefont {Aguilar}},
  \bibinfo {author} {\bibfnamefont {A.}~\bibnamefont {Ghosh}}, \bibinfo
  {author} {\bibfnamefont {S.}~\bibnamefont {Koohpayeh}}, \bibinfo {author}
  {\bibfnamefont {J.}~\bibnamefont {Krizan}}, \bibinfo {author} {\bibfnamefont
  {R.}~\bibnamefont {Cava}}, \bibinfo {author} {\bibfnamefont {O.}~\bibnamefont
  {Tchernyshyov}}, \bibinfo {author} {\bibfnamefont {T.}~\bibnamefont
  {McQueen}}, \ and\ \bibinfo {author} {\bibfnamefont {N.}~\bibnamefont
  {Armitage}},\ }\href@noop {} {\bibfield  {journal} {\bibinfo  {journal}
  {Phys. Rev. Lett.}\ }\textbf {\bibinfo {volume} {112}},\ \bibinfo {pages}
  {137403} (\bibinfo {year} {2014})}\BibitemShut {NoStop}%
\bibitem [{\citenamefont {Fava}\ \emph {et~al.}(2020)\citenamefont {Fava},
  \citenamefont {Coldea},\ and\ \citenamefont {Parameswaran}}]{fava2020glide}%
  \BibitemOpen
  \bibfield  {author} {\bibinfo {author} {\bibfnamefont {M.}~\bibnamefont
  {Fava}}, \bibinfo {author} {\bibfnamefont {R.}~\bibnamefont {Coldea}}, \ and\
  \bibinfo {author} {\bibfnamefont {S.}~\bibnamefont {Parameswaran}},\
  }\href@noop {} {\bibfield  {journal} {\bibinfo  {journal} {Proc. Natl. Acad.
  Sci.}\ }\textbf {\bibinfo {volume} {117}},\ \bibinfo {pages} {25219}
  (\bibinfo {year} {2020})}\BibitemShut {NoStop}%
\bibitem [{\citenamefont {Morris}\ \emph {et~al.}(2021)\citenamefont {Morris},
  \citenamefont {Desai}, \citenamefont {Viirok}, \citenamefont {H{\"u}vonen},
  \citenamefont {Nagel}, \citenamefont {Room}, \citenamefont {Krizan},
  \citenamefont {Cava}, \citenamefont {McQueen}, \citenamefont {Koohpayeh},
  \citenamefont {Kaul},\ and\ \citenamefont {Armitage}}]{morris2021duality}%
  \BibitemOpen
  \bibfield  {author} {\bibinfo {author} {\bibfnamefont {C.}~\bibnamefont
  {Morris}}, \bibinfo {author} {\bibfnamefont {N.}~\bibnamefont {Desai}},
  \bibinfo {author} {\bibfnamefont {J.}~\bibnamefont {Viirok}}, \bibinfo
  {author} {\bibfnamefont {D.}~\bibnamefont {H{\"u}vonen}}, \bibinfo {author}
  {\bibfnamefont {U.}~\bibnamefont {Nagel}}, \bibinfo {author} {\bibfnamefont
  {T.}~\bibnamefont {Room}}, \bibinfo {author} {\bibfnamefont {J.}~\bibnamefont
  {Krizan}}, \bibinfo {author} {\bibfnamefont {R.}~\bibnamefont {Cava}},
  \bibinfo {author} {\bibfnamefont {T.}~\bibnamefont {McQueen}}, \bibinfo
  {author} {\bibfnamefont {S.}~\bibnamefont {Koohpayeh}}, \bibinfo {author}
  {\bibfnamefont {R.~K.}\ \bibnamefont {Kaul}}, \ and\ \bibinfo {author}
  {\bibfnamefont {N.~P.}\ \bibnamefont {Armitage}},\ }\href@noop {} {\bibfield
  {journal} {\bibinfo  {journal} {Nat. Phys.}\ }\textbf {\bibinfo {volume}
  {17}},\ \bibinfo {pages} {832} (\bibinfo {year} {2021})}\BibitemShut
  {NoStop}%
\bibitem [{\citenamefont {Lefran\ifmmode~\mbox{\c{c}}\else \c{c}\fi{}ois}\
  \emph {et~al.}(2016)\citenamefont {Lefran\ifmmode~\mbox{\c{c}}\else
  \c{c}\fi{}ois}, \citenamefont {Songvilay}, \citenamefont {Robert},
  \citenamefont {Nataf}, \citenamefont {Jordan}, \citenamefont {Chaix},
  \citenamefont {Colin}, \citenamefont {Lejay}, \citenamefont {Hadj-Azzem},
  \citenamefont {Ballou},\ and\ \citenamefont {Simonet}}]{PhysRevB.94.214416}%
  \BibitemOpen
  \bibfield  {author} {\bibinfo {author} {\bibfnamefont {E.}~\bibnamefont
  {Lefran\ifmmode~\mbox{\c{c}}\else \c{c}\fi{}ois}}, \bibinfo {author}
  {\bibfnamefont {M.}~\bibnamefont {Songvilay}}, \bibinfo {author}
  {\bibfnamefont {J.}~\bibnamefont {Robert}}, \bibinfo {author} {\bibfnamefont
  {G.}~\bibnamefont {Nataf}}, \bibinfo {author} {\bibfnamefont
  {E.}~\bibnamefont {Jordan}}, \bibinfo {author} {\bibfnamefont
  {L.}~\bibnamefont {Chaix}}, \bibinfo {author} {\bibfnamefont {C.~V.}\
  \bibnamefont {Colin}}, \bibinfo {author} {\bibfnamefont {P.}~\bibnamefont
  {Lejay}}, \bibinfo {author} {\bibfnamefont {A.}~\bibnamefont {Hadj-Azzem}},
  \bibinfo {author} {\bibfnamefont {R.}~\bibnamefont {Ballou}}, \ and\ \bibinfo
  {author} {\bibfnamefont {V.}~\bibnamefont {Simonet}},\ }\href {\doibase
  10.1103/PhysRevB.94.214416} {\bibfield  {journal} {\bibinfo  {journal} {Phys.
  Rev. B}\ }\textbf {\bibinfo {volume} {94}},\ \bibinfo {pages} {214416}
  (\bibinfo {year} {2016})}\BibitemShut {NoStop}%
\bibitem [{\citenamefont {Bera}\ \emph {et~al.}(2017)\citenamefont {Bera},
  \citenamefont {Yusuf}, \citenamefont {Kumar},\ and\ \citenamefont
  {Ritter}}]{PhysRevB.95.094424}%
  \BibitemOpen
  \bibfield  {author} {\bibinfo {author} {\bibfnamefont {A.~K.}\ \bibnamefont
  {Bera}}, \bibinfo {author} {\bibfnamefont {S.~M.}\ \bibnamefont {Yusuf}},
  \bibinfo {author} {\bibfnamefont {A.}~\bibnamefont {Kumar}}, \ and\ \bibinfo
  {author} {\bibfnamefont {C.}~\bibnamefont {Ritter}},\ }\href {\doibase
  10.1103/PhysRevB.95.094424} {\bibfield  {journal} {\bibinfo  {journal} {Phys.
  Rev. B}\ }\textbf {\bibinfo {volume} {95}},\ \bibinfo {pages} {094424}
  (\bibinfo {year} {2017})}\BibitemShut {NoStop}%
\bibitem [{\citenamefont {Wong}\ \emph {et~al.}(2016)\citenamefont {Wong},
  \citenamefont {Avdeev},\ and\ \citenamefont {Ling}}]{wong2016zig}%
  \BibitemOpen
  \bibfield  {author} {\bibinfo {author} {\bibfnamefont {C.}~\bibnamefont
  {Wong}}, \bibinfo {author} {\bibfnamefont {M.}~\bibnamefont {Avdeev}}, \ and\
  \bibinfo {author} {\bibfnamefont {C.~D.}\ \bibnamefont {Ling}},\ }\href@noop
  {} {\bibfield  {journal} {\bibinfo  {journal} {J. Solid State Chem.}\
  }\textbf {\bibinfo {volume} {243}},\ \bibinfo {pages} {18} (\bibinfo {year}
  {2016})}\BibitemShut {NoStop}%
\bibitem [{\citenamefont {Chen}\ \emph {et~al.}(2021)\citenamefont {Chen},
  \citenamefont {Li}, \citenamefont {Hu}, \citenamefont {Hu}, \citenamefont
  {Yue}, \citenamefont {Sutarto}, \citenamefont {He}, \citenamefont {Iida},
  \citenamefont {Kamazawa}, \citenamefont {Yu}, \citenamefont {Lin},\ and\
  \citenamefont {Li}}]{chen2021spin}%
  \BibitemOpen
  \bibfield  {author} {\bibinfo {author} {\bibfnamefont {W.}~\bibnamefont
  {Chen}}, \bibinfo {author} {\bibfnamefont {X.}~\bibnamefont {Li}}, \bibinfo
  {author} {\bibfnamefont {Z.}~\bibnamefont {Hu}}, \bibinfo {author}
  {\bibfnamefont {Z.}~\bibnamefont {Hu}}, \bibinfo {author} {\bibfnamefont
  {L.}~\bibnamefont {Yue}}, \bibinfo {author} {\bibfnamefont {R.}~\bibnamefont
  {Sutarto}}, \bibinfo {author} {\bibfnamefont {F.}~\bibnamefont {He}},
  \bibinfo {author} {\bibfnamefont {K.}~\bibnamefont {Iida}}, \bibinfo {author}
  {\bibfnamefont {K.}~\bibnamefont {Kamazawa}}, \bibinfo {author}
  {\bibfnamefont {W.}~\bibnamefont {Yu}}, \bibinfo {author} {\bibfnamefont
  {X.}~\bibnamefont {Lin}}, \ and\ \bibinfo {author} {\bibfnamefont
  {Y.}~\bibnamefont {Li}},\ }\href {\doibase 10.1103/PhysRevB.103.L180404}
  {\bibfield  {journal} {\bibinfo  {journal} {Phys. Rev. B}\ }\textbf {\bibinfo
  {volume} {103}},\ \bibinfo {pages} {L180404} (\bibinfo {year}
  {2021})}\BibitemShut {NoStop}%
\bibitem [{\citenamefont {Chaloupka}\ \emph {et~al.}(2013)\citenamefont
  {Chaloupka}, \citenamefont {Jackeli},\ and\ \citenamefont
  {Khaliullin}}]{chaloupka2013zigzag}%
  \BibitemOpen
  \bibfield  {author} {\bibinfo {author} {\bibfnamefont {J.}~\bibnamefont
  {Chaloupka}}, \bibinfo {author} {\bibfnamefont {G.}~\bibnamefont {Jackeli}},
  \ and\ \bibinfo {author} {\bibfnamefont {G.}~\bibnamefont {Khaliullin}},\
  }\href@noop {} {\bibfield  {journal} {\bibinfo  {journal} {Phys. Rev. Lett.}\
  }\textbf {\bibinfo {volume} {110}},\ \bibinfo {pages} {097204} (\bibinfo
  {year} {2013})}\BibitemShut {NoStop}%
\bibitem [{\citenamefont {Fouet}\ \emph {et~al.}(2001)\citenamefont {Fouet},
  \citenamefont {Sindzingre},\ and\ \citenamefont
  {Lhuillier}}]{fouet2001investigation}%
  \BibitemOpen
  \bibfield  {author} {\bibinfo {author} {\bibfnamefont {J.}~\bibnamefont
  {Fouet}}, \bibinfo {author} {\bibfnamefont {P.}~\bibnamefont {Sindzingre}}, \
  and\ \bibinfo {author} {\bibfnamefont {C.}~\bibnamefont {Lhuillier}},\
  }\href@noop {} {\bibfield  {journal} {\bibinfo  {journal} {Eur. Phys. J. B}\
  }\textbf {\bibinfo {volume} {20}},\ \bibinfo {pages} {241} (\bibinfo {year}
  {2001})}\BibitemShut {NoStop}%
\bibitem [{\citenamefont {Kimchi}\ and\ \citenamefont
  {You}(2011)}]{kimchi2011kitaev}%
  \BibitemOpen
  \bibfield  {author} {\bibinfo {author} {\bibfnamefont {I.}~\bibnamefont
  {Kimchi}}\ and\ \bibinfo {author} {\bibfnamefont {Y.-Z.}\ \bibnamefont
  {You}},\ }\href@noop {} {\bibfield  {journal} {\bibinfo  {journal} {Phys.
  Rev. B}\ }\textbf {\bibinfo {volume} {84}},\ \bibinfo {pages} {180407}
  (\bibinfo {year} {2011})}\BibitemShut {NoStop}%
\bibitem [{\citenamefont {Songvilay}\ \emph {et~al.}(2020)\citenamefont
  {Songvilay}, \citenamefont {Robert}, \citenamefont {Petit}, \citenamefont
  {Rodriguez-Rivera}, \citenamefont {Ratcliff}, \citenamefont {Damay},
  \citenamefont {Bal\'edent}, \citenamefont {Jim\'enez-Ruiz}, \citenamefont
  {Lejay}, \citenamefont {Pachoud}, \citenamefont {Hadj-Azzem}, \citenamefont
  {Simonet},\ and\ \citenamefont {Stock}}]{songvilay2020kitaev}%
  \BibitemOpen
  \bibfield  {author} {\bibinfo {author} {\bibfnamefont {M.}~\bibnamefont
  {Songvilay}}, \bibinfo {author} {\bibfnamefont {J.}~\bibnamefont {Robert}},
  \bibinfo {author} {\bibfnamefont {S.}~\bibnamefont {Petit}}, \bibinfo
  {author} {\bibfnamefont {J.~A.}\ \bibnamefont {Rodriguez-Rivera}}, \bibinfo
  {author} {\bibfnamefont {W.~D.}\ \bibnamefont {Ratcliff}}, \bibinfo {author}
  {\bibfnamefont {F.}~\bibnamefont {Damay}}, \bibinfo {author} {\bibfnamefont
  {V.}~\bibnamefont {Bal\'edent}}, \bibinfo {author} {\bibfnamefont
  {M.}~\bibnamefont {Jim\'enez-Ruiz}}, \bibinfo {author} {\bibfnamefont
  {P.}~\bibnamefont {Lejay}}, \bibinfo {author} {\bibfnamefont
  {E.}~\bibnamefont {Pachoud}}, \bibinfo {author} {\bibfnamefont
  {A.}~\bibnamefont {Hadj-Azzem}}, \bibinfo {author} {\bibfnamefont
  {V.}~\bibnamefont {Simonet}}, \ and\ \bibinfo {author} {\bibfnamefont
  {C.}~\bibnamefont {Stock}},\ }\href {\doibase 10.1103/PhysRevB.102.224429}
  {\bibfield  {journal} {\bibinfo  {journal} {Phys. Rev. B}\ }\textbf {\bibinfo
  {volume} {102}},\ \bibinfo {pages} {224429} (\bibinfo {year}
  {2020})}\BibitemShut {NoStop}%
\bibitem [{\citenamefont {Kim}\ \emph {et~al.}(2021)\citenamefont {Kim},
  \citenamefont {Jeong}, \citenamefont {Lin}, \citenamefont {Park},
  \citenamefont {Masuda}, \citenamefont {Asai}, \citenamefont {Itoh},
  \citenamefont {Kim}, \citenamefont {Zhou}, \citenamefont {Ma},\ and\
  \citenamefont {Park}}]{kim2021antiferromagnetic}%
  \BibitemOpen
  \bibfield  {author} {\bibinfo {author} {\bibfnamefont {C.}~\bibnamefont
  {Kim}}, \bibinfo {author} {\bibfnamefont {J.}~\bibnamefont {Jeong}}, \bibinfo
  {author} {\bibfnamefont {G.}~\bibnamefont {Lin}}, \bibinfo {author}
  {\bibfnamefont {P.}~\bibnamefont {Park}}, \bibinfo {author} {\bibfnamefont
  {T.}~\bibnamefont {Masuda}}, \bibinfo {author} {\bibfnamefont
  {S.}~\bibnamefont {Asai}}, \bibinfo {author} {\bibfnamefont {S.}~\bibnamefont
  {Itoh}}, \bibinfo {author} {\bibfnamefont {H.-S.}\ \bibnamefont {Kim}},
  \bibinfo {author} {\bibfnamefont {H.}~\bibnamefont {Zhou}}, \bibinfo {author}
  {\bibfnamefont {J.}~\bibnamefont {Ma}}, \ and\ \bibinfo {author}
  {\bibfnamefont {J.-G.}\ \bibnamefont {Park}},\ }\href@noop {} {\bibfield
  {journal} {\bibinfo  {journal} {J. Phys. Condens. Matter}\ }\textbf {\bibinfo
  {volume} {34}},\ \bibinfo {pages} {045802} (\bibinfo {year}
  {2021})}\BibitemShut {NoStop}%
\bibitem [{\citenamefont {Lin}\ \emph {et~al.}(2021)\citenamefont {Lin},
  \citenamefont {Jeong}, \citenamefont {Kim}, \citenamefont {Wang},
  \citenamefont {Huang}, \citenamefont {Masuda}, \citenamefont {Asai},
  \citenamefont {Itoh}, \citenamefont {G{\"u}nther}, \citenamefont {Russina},
  \citenamefont {Lu}, \citenamefont {Sheng}, \citenamefont {Wang},
  \citenamefont {Wang}, \citenamefont {Wang}, \citenamefont {Ren},
  \citenamefont {Xi}, \citenamefont {Tong}, \citenamefont {Ling}, \citenamefont
  {Liu}, \citenamefont {Mei}, \citenamefont {Qu}, \citenamefont {Zhou},
  \citenamefont {Park}, \citenamefont {Wan},\ and\ \citenamefont
  {Ma}}]{lin2021field}%
  \BibitemOpen
  \bibfield  {author} {\bibinfo {author} {\bibfnamefont {G.}~\bibnamefont
  {Lin}}, \bibinfo {author} {\bibfnamefont {J.}~\bibnamefont {Jeong}}, \bibinfo
  {author} {\bibfnamefont {C.}~\bibnamefont {Kim}}, \bibinfo {author}
  {\bibfnamefont {Y.}~\bibnamefont {Wang}}, \bibinfo {author} {\bibfnamefont
  {Q.}~\bibnamefont {Huang}}, \bibinfo {author} {\bibfnamefont
  {T.}~\bibnamefont {Masuda}}, \bibinfo {author} {\bibfnamefont
  {S.}~\bibnamefont {Asai}}, \bibinfo {author} {\bibfnamefont {S.}~\bibnamefont
  {Itoh}}, \bibinfo {author} {\bibfnamefont {G.}~\bibnamefont {G{\"u}nther}},
  \bibinfo {author} {\bibfnamefont {M.}~\bibnamefont {Russina}}, \bibinfo
  {author} {\bibfnamefont {Z.}~\bibnamefont {Lu}}, \bibinfo {author}
  {\bibfnamefont {J.}~\bibnamefont {Sheng}}, \bibinfo {author} {\bibfnamefont
  {L.}~\bibnamefont {Wang}}, \bibinfo {author} {\bibfnamefont {J.}~\bibnamefont
  {Wang}}, \bibinfo {author} {\bibfnamefont {G.}~\bibnamefont {Wang}}, \bibinfo
  {author} {\bibfnamefont {Q.}~\bibnamefont {Ren}}, \bibinfo {author}
  {\bibfnamefont {C.}~\bibnamefont {Xi}}, \bibinfo {author} {\bibfnamefont
  {W.}~\bibnamefont {Tong}}, \bibinfo {author} {\bibfnamefont {L.}~\bibnamefont
  {Ling}}, \bibinfo {author} {\bibfnamefont {Z.}~\bibnamefont {Liu}}, \bibinfo
  {author} {\bibfnamefont {J.}~\bibnamefont {Mei}}, \bibinfo {author}
  {\bibfnamefont {Z.}~\bibnamefont {Qu}}, \bibinfo {author} {\bibfnamefont
  {H.}~\bibnamefont {Zhou}}, \bibinfo {author} {\bibfnamefont {J.-G.}\
  \bibnamefont {Park}}, \bibinfo {author} {\bibfnamefont {Y.}~\bibnamefont
  {Wan}}, \ and\ \bibinfo {author} {\bibfnamefont {J.}~\bibnamefont {Ma}},\
  }\href@noop {} {\bibfield  {journal} {\bibinfo  {journal} {Nat. Commun.}\
  }\textbf {\bibinfo {volume} {12}},\ \bibinfo {pages} {1} (\bibinfo {year}
  {2021})}\BibitemShut {NoStop}%
\bibitem [{\citenamefont {Nair}\ \emph {et~al.}(2018)\citenamefont {Nair},
  \citenamefont {Brown}, \citenamefont {Coldren}, \citenamefont {Hester},
  \citenamefont {Gelfand}, \citenamefont {Podlesnyak}, \citenamefont {Huang},\
  and\ \citenamefont {Ross}}]{nair2018short}%
  \BibitemOpen
  \bibfield  {author} {\bibinfo {author} {\bibfnamefont {H.~S.}\ \bibnamefont
  {Nair}}, \bibinfo {author} {\bibfnamefont {J.}~\bibnamefont {Brown}},
  \bibinfo {author} {\bibfnamefont {E.}~\bibnamefont {Coldren}}, \bibinfo
  {author} {\bibfnamefont {G.}~\bibnamefont {Hester}}, \bibinfo {author}
  {\bibfnamefont {M.}~\bibnamefont {Gelfand}}, \bibinfo {author} {\bibfnamefont
  {A.}~\bibnamefont {Podlesnyak}}, \bibinfo {author} {\bibfnamefont
  {Q.}~\bibnamefont {Huang}}, \ and\ \bibinfo {author} {\bibfnamefont
  {K.}~\bibnamefont {Ross}},\ }\href@noop {} {\bibfield  {journal} {\bibinfo
  {journal} {Phys. Rev. B}\ }\textbf {\bibinfo {volume} {97}},\ \bibinfo
  {pages} {134409} (\bibinfo {year} {2018})}\BibitemShut {NoStop}%
\bibitem [{\citenamefont {Regnault}\ \emph {et~al.}(2006)\citenamefont
  {Regnault}, \citenamefont {Boullier},\ and\ \citenamefont
  {Henry}}]{regnault2006investigation}%
  \BibitemOpen
  \bibfield  {author} {\bibinfo {author} {\bibfnamefont {L.}~\bibnamefont
  {Regnault}}, \bibinfo {author} {\bibfnamefont {C.}~\bibnamefont {Boullier}},
  \ and\ \bibinfo {author} {\bibfnamefont {J.}~\bibnamefont {Henry}},\
  }\href@noop {} {\bibfield  {journal} {\bibinfo  {journal} {Physica B Condens.
  Matter}\ }\textbf {\bibinfo {volume} {385}},\ \bibinfo {pages} {425}
  (\bibinfo {year} {2006})}\BibitemShut {NoStop}%
\bibitem [{\citenamefont {Regnault}\ \emph {et~al.}(2018)\citenamefont
  {Regnault}, \citenamefont {Boullier},\ and\ \citenamefont
  {Lorenzo}}]{regnault2018polarized}%
  \BibitemOpen
  \bibfield  {author} {\bibinfo {author} {\bibfnamefont {L.-P.}\ \bibnamefont
  {Regnault}}, \bibinfo {author} {\bibfnamefont {C.}~\bibnamefont {Boullier}},
  \ and\ \bibinfo {author} {\bibfnamefont {J.}~\bibnamefont {Lorenzo}},\
  }\href@noop {} {\bibfield  {journal} {\bibinfo  {journal} {Heliyon}\ }\textbf
  {\bibinfo {volume} {4}},\ \bibinfo {pages} {e00507} (\bibinfo {year}
  {2018})}\BibitemShut {NoStop}%
\bibitem [{\citenamefont {Regnault}\ \emph {et~al.}(1977)\citenamefont
  {Regnault}, \citenamefont {Burlet},\ and\ \citenamefont
  {Rossat-Mignod}}]{regnault1977magnetic}%
  \BibitemOpen
  \bibfield  {author} {\bibinfo {author} {\bibfnamefont {L.}~\bibnamefont
  {Regnault}}, \bibinfo {author} {\bibfnamefont {P.}~\bibnamefont {Burlet}}, \
  and\ \bibinfo {author} {\bibfnamefont {J.}~\bibnamefont {Rossat-Mignod}},\
  }\href@noop {} {\bibfield  {journal} {\bibinfo  {journal} {Physica B Condens.
  Matter}\ }\textbf {\bibinfo {volume} {86}},\ \bibinfo {pages} {660} (\bibinfo
  {year} {1977})}\BibitemShut {NoStop}%
\bibitem [{\citenamefont {Zhong}\ \emph {et~al.}(2020)\citenamefont {Zhong},
  \citenamefont {Gao}, \citenamefont {Ong},\ and\ \citenamefont
  {Cava}}]{zhong2020weak}%
  \BibitemOpen
  \bibfield  {author} {\bibinfo {author} {\bibfnamefont {R.}~\bibnamefont
  {Zhong}}, \bibinfo {author} {\bibfnamefont {T.}~\bibnamefont {Gao}}, \bibinfo
  {author} {\bibfnamefont {N.~P.}\ \bibnamefont {Ong}}, \ and\ \bibinfo
  {author} {\bibfnamefont {R.~J.}\ \bibnamefont {Cava}},\ }\href@noop {}
  {\bibfield  {journal} {\bibinfo  {journal} {Sci. Adv.}\ }\textbf {\bibinfo
  {volume} {6}},\ \bibinfo {pages} {eaay6953} (\bibinfo {year}
  {2020})}\BibitemShut {NoStop}%
\bibitem [{\citenamefont {Zhang}\ \emph {et~al.}(2021)\citenamefont {Zhang},
  \citenamefont {Xu}, \citenamefont {Zhong}, \citenamefont {Cava},
  \citenamefont {Drichko},\ and\ \citenamefont {Armitage}}]{zhang2021and}%
  \BibitemOpen
  \bibfield  {author} {\bibinfo {author} {\bibfnamefont {X.}~\bibnamefont
  {Zhang}}, \bibinfo {author} {\bibfnamefont {Y.}~\bibnamefont {Xu}}, \bibinfo
  {author} {\bibfnamefont {R.}~\bibnamefont {Zhong}}, \bibinfo {author}
  {\bibfnamefont {R.}~\bibnamefont {Cava}}, \bibinfo {author} {\bibfnamefont
  {N.}~\bibnamefont {Drichko}}, \ and\ \bibinfo {author} {\bibfnamefont
  {N.}~\bibnamefont {Armitage}},\ }\href@noop {} {\bibfield  {journal}
  {\bibinfo  {journal} {arXiv preprint arXiv:2106.13418}\ } (\bibinfo {year}
  {2021})}\BibitemShut {NoStop}%
\bibitem [{\citenamefont {Shi}\ \emph {et~al.}(2021)\citenamefont {Shi},
  \citenamefont {Wang}, \citenamefont {Zhong}, \citenamefont {Wang},
  \citenamefont {Hu}, \citenamefont {Zhang}, \citenamefont {Liu}, \citenamefont
  {Dong}, \citenamefont {Wang},\ and\ \citenamefont {Wang}}]{shi2021magnetic}%
  \BibitemOpen
  \bibfield  {author} {\bibinfo {author} {\bibfnamefont {L.}~\bibnamefont
  {Shi}}, \bibinfo {author} {\bibfnamefont {X.}~\bibnamefont {Wang}}, \bibinfo
  {author} {\bibfnamefont {R.}~\bibnamefont {Zhong}}, \bibinfo {author}
  {\bibfnamefont {Z.}~\bibnamefont {Wang}}, \bibinfo {author} {\bibfnamefont
  {T.}~\bibnamefont {Hu}}, \bibinfo {author} {\bibfnamefont {S.}~\bibnamefont
  {Zhang}}, \bibinfo {author} {\bibfnamefont {Q.}~\bibnamefont {Liu}}, \bibinfo
  {author} {\bibfnamefont {T.}~\bibnamefont {Dong}}, \bibinfo {author}
  {\bibfnamefont {F.}~\bibnamefont {Wang}}, \ and\ \bibinfo {author}
  {\bibfnamefont {N.}~\bibnamefont {Wang}},\ }\href@noop {} {\bibfield
  {journal} {\bibinfo  {journal} {Phys. Rev. B}\ }\textbf {\bibinfo {volume}
  {104}},\ \bibinfo {pages} {144408} (\bibinfo {year} {2021})}\BibitemShut
  {NoStop}%
\bibitem [{\citenamefont {Das}\ \emph {et~al.}(2021)\citenamefont {Das},
  \citenamefont {Voleti}, \citenamefont {Saha-Dasgupta},\ and\ \citenamefont
  {Paramekanti}}]{das2021xy}%
  \BibitemOpen
  \bibfield  {author} {\bibinfo {author} {\bibfnamefont {S.}~\bibnamefont
  {Das}}, \bibinfo {author} {\bibfnamefont {S.}~\bibnamefont {Voleti}},
  \bibinfo {author} {\bibfnamefont {T.}~\bibnamefont {Saha-Dasgupta}}, \ and\
  \bibinfo {author} {\bibfnamefont {A.}~\bibnamefont {Paramekanti}},\
  }\href@noop {} {\bibfield  {journal} {\bibinfo  {journal} {Phys. Rev. B}\
  }\textbf {\bibinfo {volume} {104}},\ \bibinfo {pages} {134425} (\bibinfo
  {year} {2021})}\BibitemShut {NoStop}%
\bibitem [{\citenamefont {Maksimov}\ \emph {et~al.}(2022)\citenamefont
  {Maksimov}, \citenamefont {Ushakov}, \citenamefont {Pchelkina}, \citenamefont
  {Li}, \citenamefont {Winter},\ and\ \citenamefont
  {Streltsov}}]{maksimov2022}%
  \BibitemOpen
  \bibfield  {author} {\bibinfo {author} {\bibfnamefont {P.~A.}\ \bibnamefont
  {Maksimov}}, \bibinfo {author} {\bibfnamefont {A.~V.}\ \bibnamefont
  {Ushakov}}, \bibinfo {author} {\bibfnamefont {Z.~V.}\ \bibnamefont
  {Pchelkina}}, \bibinfo {author} {\bibfnamefont {Y.}~\bibnamefont {Li}},
  \bibinfo {author} {\bibfnamefont {S.~M.}\ \bibnamefont {Winter}}, \ and\
  \bibinfo {author} {\bibfnamefont {S.~V.}\ \bibnamefont {Streltsov}},\
  }\href@noop {} {\bibfield  {journal} {\bibinfo  {journal} {arXiv preprint
  arXiv:2204.09695}\ } (\bibinfo {year} {2022})}\BibitemShut {NoStop}%
\bibitem [{Note1()}]{Note1}%
  \BibitemOpen
  \bibinfo {note} {The coefficients $U_{\alpha \beta \gamma \delta }$ may be
  grouped according to the number of unique orbital indices, from one to four.
  For example, the intra-orbital Hubbard terms $n_{i,\alpha ,\uparrow
  }n_{i,\alpha ,\downarrow }$ have one unique index $\alpha $, while the
  inter-orbital Hubbard terms $n_{i,\alpha ,\sigma }n_{i,\beta ,\sigma ^\prime
  }$ have two unique indices $\alpha ,\beta $. In the spherically symmetric
  approximation \cite {sugano2012multiplets}, the Coulomb coefficients with
  three and four indices vanish unless at least one of the orbitals is an $e_g$
  orbital. For this reason, $t_{2g}$-only (and $e_g$-only) models reduce to the
  familiar Kanamori form\cite {georges2013strong,pavarini2014dmft}, which
  includes only Hubbard density-density repulsion, Hund's exchange, and
  pair-hopping contributions. However, when both $e_g$ and $t_{2g}$ orbitals
  are considered together, it is important to include the full rotationally
  symmetric Coulomb terms. This is particularly true when computing anisotropic
  magnetic exchange, because any approximations to the Coulomb Hamiltonian are
  likely to explicitly break rotational symmetry, leading to erroneous sources
  of anisotropy.}\BibitemShut {Stop}%
\bibitem [{\citenamefont {Sugano}(2012)}]{sugano2012multiplets}%
  \BibitemOpen
  \bibfield  {author} {\bibinfo {author} {\bibfnamefont {S.}~\bibnamefont
  {Sugano}},\ }\href@noop {} {\emph {\bibinfo {title} {Multiplets of
  transition-metal ions in crystals}}}\ (\bibinfo  {publisher} {Elsevier},\
  \bibinfo {year} {2012})\BibitemShut {NoStop}%
\bibitem [{\citenamefont {Pavarini}(2014)}]{pavarini2014dmft}%
  \BibitemOpen
  \bibfield  {author} {\bibinfo {author} {\bibfnamefont {E.}~\bibnamefont
  {Pavarini}},\ }in\ \href@noop {} {\emph {\bibinfo {booktitle} {Many-Electron
  Approaches in Physics, Chemistry and Mathematics}}}\ (\bibinfo  {publisher}
  {Springer},\ \bibinfo {year} {2014})\ pp.\ \bibinfo {pages}
  {321--341}\BibitemShut {NoStop}%
\bibitem [{\citenamefont {Sarte}\ \emph {et~al.}(2018)\citenamefont {Sarte},
  \citenamefont {Cowley}, \citenamefont {Rodriguez}, \citenamefont {Pachoud},
  \citenamefont {Le}, \citenamefont {Garc\'{\i}a-Sakai}, \citenamefont
  {Taylor}, \citenamefont {Frost}, \citenamefont {Prabhakaran}, \citenamefont
  {MacEwen}, \citenamefont {Kitada}, \citenamefont {Browne}, \citenamefont
  {Songvilay}, \citenamefont {Yamani}, \citenamefont {Buyers}, \citenamefont
  {Attfield},\ and\ \citenamefont {Stock}}]{PhysRevB.98.024415}%
  \BibitemOpen
  \bibfield  {author} {\bibinfo {author} {\bibfnamefont {P.~M.}\ \bibnamefont
  {Sarte}}, \bibinfo {author} {\bibfnamefont {R.~A.}\ \bibnamefont {Cowley}},
  \bibinfo {author} {\bibfnamefont {E.~E.}\ \bibnamefont {Rodriguez}}, \bibinfo
  {author} {\bibfnamefont {E.}~\bibnamefont {Pachoud}}, \bibinfo {author}
  {\bibfnamefont {D.}~\bibnamefont {Le}}, \bibinfo {author} {\bibfnamefont
  {V.}~\bibnamefont {Garc\'{\i}a-Sakai}}, \bibinfo {author} {\bibfnamefont
  {J.~W.}\ \bibnamefont {Taylor}}, \bibinfo {author} {\bibfnamefont {C.~D.}\
  \bibnamefont {Frost}}, \bibinfo {author} {\bibfnamefont {D.}~\bibnamefont
  {Prabhakaran}}, \bibinfo {author} {\bibfnamefont {C.}~\bibnamefont
  {MacEwen}}, \bibinfo {author} {\bibfnamefont {A.}~\bibnamefont {Kitada}},
  \bibinfo {author} {\bibfnamefont {A.~J.}\ \bibnamefont {Browne}}, \bibinfo
  {author} {\bibfnamefont {M.}~\bibnamefont {Songvilay}}, \bibinfo {author}
  {\bibfnamefont {Z.}~\bibnamefont {Yamani}}, \bibinfo {author} {\bibfnamefont
  {W.~J.~L.}\ \bibnamefont {Buyers}}, \bibinfo {author} {\bibfnamefont {J.~P.}\
  \bibnamefont {Attfield}}, \ and\ \bibinfo {author} {\bibfnamefont
  {C.}~\bibnamefont {Stock}},\ }\href {\doibase 10.1103/PhysRevB.98.024415}
  {\bibfield  {journal} {\bibinfo  {journal} {Phys. Rev. B}\ }\textbf {\bibinfo
  {volume} {98}},\ \bibinfo {pages} {024415} (\bibinfo {year}
  {2018})}\BibitemShut {NoStop}%
\bibitem [{\citenamefont {Ross}\ \emph {et~al.}(2017)\citenamefont {Ross},
  \citenamefont {Brown}, \citenamefont {Cava}, \citenamefont {Krizan},
  \citenamefont {Nagler}, \citenamefont {Rodriguez-Rivera},\ and\ \citenamefont
  {Stone}}]{ross2017single}%
  \BibitemOpen
  \bibfield  {author} {\bibinfo {author} {\bibfnamefont {K.~A.}\ \bibnamefont
  {Ross}}, \bibinfo {author} {\bibfnamefont {J.}~\bibnamefont {Brown}},
  \bibinfo {author} {\bibfnamefont {R.}~\bibnamefont {Cava}}, \bibinfo {author}
  {\bibfnamefont {J.}~\bibnamefont {Krizan}}, \bibinfo {author} {\bibfnamefont
  {S.~E.}\ \bibnamefont {Nagler}}, \bibinfo {author} {\bibfnamefont
  {J.}~\bibnamefont {Rodriguez-Rivera}}, \ and\ \bibinfo {author}
  {\bibfnamefont {M.~B.}\ \bibnamefont {Stone}},\ }\href@noop {} {\bibfield
  {journal} {\bibinfo  {journal} {Phys. Rev. B}\ }\textbf {\bibinfo {volume}
  {95}},\ \bibinfo {pages} {144414} (\bibinfo {year} {2017})}\BibitemShut
  {NoStop}%
\bibitem [{\citenamefont {Koepernik}\ and\ \citenamefont
  {Eschrig}(1999)}]{koepernik1999full}%
  \BibitemOpen
  \bibfield  {author} {\bibinfo {author} {\bibfnamefont {K.}~\bibnamefont
  {Koepernik}}\ and\ \bibinfo {author} {\bibfnamefont {H.}~\bibnamefont
  {Eschrig}},\ }\href@noop {} {\bibfield  {journal} {\bibinfo  {journal}
  {Physical Review B}\ }\textbf {\bibinfo {volume} {59}},\ \bibinfo {pages}
  {1743} (\bibinfo {year} {1999})}\BibitemShut {NoStop}%
\bibitem [{\citenamefont {Opahle}\ \emph {et~al.}(1999)\citenamefont {Opahle},
  \citenamefont {Koepernik},\ and\ \citenamefont {Eschrig}}]{opahle1999full}%
  \BibitemOpen
  \bibfield  {author} {\bibinfo {author} {\bibfnamefont {I.}~\bibnamefont
  {Opahle}}, \bibinfo {author} {\bibfnamefont {K.}~\bibnamefont {Koepernik}}, \
  and\ \bibinfo {author} {\bibfnamefont {H.}~\bibnamefont {Eschrig}},\
  }\href@noop {} {\bibfield  {journal} {\bibinfo  {journal} {Physical Review
  B}\ }\textbf {\bibinfo {volume} {60}},\ \bibinfo {pages} {14035} (\bibinfo
  {year} {1999})}\BibitemShut {NoStop}%
\bibitem [{\citenamefont {Perdew}\ \emph {et~al.}(1996)\citenamefont {Perdew},
  \citenamefont {Burke},\ and\ \citenamefont
  {Ernzerhof}}]{perdew1996generalized}%
  \BibitemOpen
  \bibfield  {author} {\bibinfo {author} {\bibfnamefont {J.~P.}\ \bibnamefont
  {Perdew}}, \bibinfo {author} {\bibfnamefont {K.}~\bibnamefont {Burke}}, \
  and\ \bibinfo {author} {\bibfnamefont {M.}~\bibnamefont {Ernzerhof}},\
  }\href@noop {} {\bibfield  {journal} {\bibinfo  {journal} {Physical review
  letters}\ }\textbf {\bibinfo {volume} {77}},\ \bibinfo {pages} {3865}
  (\bibinfo {year} {1996})}\BibitemShut {NoStop}%
\bibitem [{\citenamefont {Gray}(1965)}]{gray1965electrons}%
  \BibitemOpen
  \bibfield  {author} {\bibinfo {author} {\bibfnamefont {H.~B.}\ \bibnamefont
  {Gray}},\ }\href@noop {} {\emph {\bibinfo {title} {Electrons and chemical
  bonding}}}\ (\bibinfo  {publisher} {WA Benjamin, Inc.},\ \bibinfo {year}
  {1965})\BibitemShut {NoStop}%
\bibitem [{\citenamefont {Sarvezuk}\ \emph {et~al.}(2011)\citenamefont
  {Sarvezuk}, \citenamefont {Kinast}, \citenamefont {Colin}, \citenamefont
  {Gusm{\~a}o}, \citenamefont {Da~Cunha},\ and\ \citenamefont
  {Isnard}}]{sarvezuk2011new}%
  \BibitemOpen
  \bibfield  {author} {\bibinfo {author} {\bibfnamefont {P.~W.~C.}\
  \bibnamefont {Sarvezuk}}, \bibinfo {author} {\bibfnamefont {E.~J.}\
  \bibnamefont {Kinast}}, \bibinfo {author} {\bibfnamefont {C.}~\bibnamefont
  {Colin}}, \bibinfo {author} {\bibfnamefont {M.}~\bibnamefont {Gusm{\~a}o}},
  \bibinfo {author} {\bibfnamefont {J.}~\bibnamefont {Da~Cunha}}, \ and\
  \bibinfo {author} {\bibfnamefont {O.}~\bibnamefont {Isnard}},\ }\href@noop {}
  {\bibfield  {journal} {\bibinfo  {journal} {Int. J. Appl. Phys.}\ }\textbf
  {\bibinfo {volume} {109}},\ \bibinfo {pages} {07E160} (\bibinfo {year}
  {2011})}\BibitemShut {NoStop}%
\bibitem [{\citenamefont {Dordevi{\'c}}(2008)}]{djordevic2008baco2}%
  \BibitemOpen
  \bibfield  {author} {\bibinfo {author} {\bibfnamefont {T.}~\bibnamefont
  {Dordevi{\'c}}},\ }\href@noop {} {\bibfield  {journal} {\bibinfo  {journal}
  {Acta Crystallogr. E}\ }\textbf {\bibinfo {volume} {64}},\ \bibinfo {pages}
  {i58} (\bibinfo {year} {2008})}\BibitemShut {NoStop}%
\bibitem [{\citenamefont {Xiao}\ \emph {et~al.}(2019)\citenamefont {Xiao},
  \citenamefont {Xia}, \citenamefont {Zhang}, \citenamefont {Yue},
  \citenamefont {Huang}, \citenamefont {Zhang}, \citenamefont {Yang},
  \citenamefont {Song}, \citenamefont {Wei}, \citenamefont {Deng},\ and\
  \citenamefont {Jiang}}]{xiao2019crystal}%
  \BibitemOpen
  \bibfield  {author} {\bibinfo {author} {\bibfnamefont {G.}~\bibnamefont
  {Xiao}}, \bibinfo {author} {\bibfnamefont {Z.}~\bibnamefont {Xia}}, \bibinfo
  {author} {\bibfnamefont {W.}~\bibnamefont {Zhang}}, \bibinfo {author}
  {\bibfnamefont {X.}~\bibnamefont {Yue}}, \bibinfo {author} {\bibfnamefont
  {S.}~\bibnamefont {Huang}}, \bibinfo {author} {\bibfnamefont
  {X.}~\bibnamefont {Zhang}}, \bibinfo {author} {\bibfnamefont
  {F.}~\bibnamefont {Yang}}, \bibinfo {author} {\bibfnamefont {Y.}~\bibnamefont
  {Song}}, \bibinfo {author} {\bibfnamefont {M.}~\bibnamefont {Wei}}, \bibinfo
  {author} {\bibfnamefont {H.}~\bibnamefont {Deng}}, \ and\ \bibinfo {author}
  {\bibfnamefont {D.}~\bibnamefont {Jiang}},\ }\href@noop {} {\bibfield
  {journal} {\bibinfo  {journal} {Crystal Growth \& Design}\ }\textbf {\bibinfo
  {volume} {19}},\ \bibinfo {pages} {2658} (\bibinfo {year}
  {2019})}\BibitemShut {NoStop}%
\bibitem [{Note2()}]{Note2}%
  \BibitemOpen
  \bibinfo {note} {The Na$_2$Co$_2$TeO$_6$ structure contains disorder in the
  Na position, in which each Na position has occupancy 2/3. To perform
  calculations, we artificially increased the occupancy to 1, which corresponds
  to Na$_3$Co$_2$TeO$_6$. It is expected this change in the filling should have
  minimal impact on the computed hoppings.}\BibitemShut {Stop}%
\bibitem [{\citenamefont {Wilkinson}\ \emph {et~al.}(1959)\citenamefont
  {Wilkinson}, \citenamefont {Cable}, \citenamefont {Wollan},\ and\
  \citenamefont {Koehler}}]{wilkinson1959neutron}%
  \BibitemOpen
  \bibfield  {author} {\bibinfo {author} {\bibfnamefont {M.}~\bibnamefont
  {Wilkinson}}, \bibinfo {author} {\bibfnamefont {J.}~\bibnamefont {Cable}},
  \bibinfo {author} {\bibfnamefont {E.}~\bibnamefont {Wollan}}, \ and\ \bibinfo
  {author} {\bibfnamefont {W.}~\bibnamefont {Koehler}},\ }\href@noop {}
  {\bibfield  {journal} {\bibinfo  {journal} {Phys. Rev.}\ }\textbf {\bibinfo
  {volume} {113}},\ \bibinfo {pages} {497} (\bibinfo {year}
  {1959})}\BibitemShut {NoStop}%
\bibitem [{\citenamefont {Wyckoff}\ and\ \citenamefont
  {Wyckoff}(1963)}]{wyckoff1963crystal}%
  \BibitemOpen
  \bibfield  {author} {\bibinfo {author} {\bibfnamefont {R.~W.~G.}\
  \bibnamefont {Wyckoff}}\ and\ \bibinfo {author} {\bibfnamefont {R.~W.}\
  \bibnamefont {Wyckoff}},\ }\href@noop {} {\emph {\bibinfo {title} {Crystal
  structures}}},\ Vol.~\bibinfo {volume} {1}\ (\bibinfo  {publisher}
  {Interscience publishers New York},\ \bibinfo {year} {1963})\BibitemShut
  {NoStop}%
\bibitem [{\citenamefont {Wellm}\ \emph {et~al.}(2021)\citenamefont {Wellm},
  \citenamefont {Roscher}, \citenamefont {Zeisner}, \citenamefont {Alfonsov},
  \citenamefont {Zhong}, \citenamefont {Cava}, \citenamefont {Savoyant},
  \citenamefont {Hayn}, \citenamefont {van~den Brink}, \citenamefont
  {B\"uchner}, \citenamefont {Janson},\ and\ \citenamefont
  {Kataev}}]{wellm2021frustration}%
  \BibitemOpen
  \bibfield  {author} {\bibinfo {author} {\bibfnamefont {C.}~\bibnamefont
  {Wellm}}, \bibinfo {author} {\bibfnamefont {W.}~\bibnamefont {Roscher}},
  \bibinfo {author} {\bibfnamefont {J.}~\bibnamefont {Zeisner}}, \bibinfo
  {author} {\bibfnamefont {A.}~\bibnamefont {Alfonsov}}, \bibinfo {author}
  {\bibfnamefont {R.}~\bibnamefont {Zhong}}, \bibinfo {author} {\bibfnamefont
  {R.~J.}\ \bibnamefont {Cava}}, \bibinfo {author} {\bibfnamefont
  {A.}~\bibnamefont {Savoyant}}, \bibinfo {author} {\bibfnamefont
  {R.}~\bibnamefont {Hayn}}, \bibinfo {author} {\bibfnamefont {J.}~\bibnamefont
  {van~den Brink}}, \bibinfo {author} {\bibfnamefont {B.}~\bibnamefont
  {B\"uchner}}, \bibinfo {author} {\bibfnamefont {O.}~\bibnamefont {Janson}}, \
  and\ \bibinfo {author} {\bibfnamefont {V.}~\bibnamefont {Kataev}},\ }\href
  {\doibase 10.1103/PhysRevB.104.L100420} {\bibfield  {journal} {\bibinfo
  {journal} {Phys. Rev. B}\ }\textbf {\bibinfo {volume} {104}},\ \bibinfo
  {pages} {L100420} (\bibinfo {year} {2021})}\BibitemShut {NoStop}%
\bibitem [{\citenamefont {Ross}\ \emph {et~al.}(2011)\citenamefont {Ross},
  \citenamefont {Savary}, \citenamefont {Gaulin},\ and\ \citenamefont
  {Balents}}]{ross2011quantum}%
  \BibitemOpen
  \bibfield  {author} {\bibinfo {author} {\bibfnamefont {K.~A.}\ \bibnamefont
  {Ross}}, \bibinfo {author} {\bibfnamefont {L.}~\bibnamefont {Savary}},
  \bibinfo {author} {\bibfnamefont {B.~D.}\ \bibnamefont {Gaulin}}, \ and\
  \bibinfo {author} {\bibfnamefont {L.}~\bibnamefont {Balents}},\ }\href@noop
  {} {\bibfield  {journal} {\bibinfo  {journal} {Phys. Rev. X}\ }\textbf
  {\bibinfo {volume} {1}},\ \bibinfo {pages} {021002} (\bibinfo {year}
  {2011})}\BibitemShut {NoStop}%
\bibitem [{\citenamefont {Maksimov}\ \emph {et~al.}(2019)\citenamefont
  {Maksimov}, \citenamefont {Zhu}, \citenamefont {White},\ and\ \citenamefont
  {Chernyshev}}]{maksimov2019anisotropic}%
  \BibitemOpen
  \bibfield  {author} {\bibinfo {author} {\bibfnamefont {P.}~\bibnamefont
  {Maksimov}}, \bibinfo {author} {\bibfnamefont {Z.}~\bibnamefont {Zhu}},
  \bibinfo {author} {\bibfnamefont {S.~R.}\ \bibnamefont {White}}, \ and\
  \bibinfo {author} {\bibfnamefont {A.}~\bibnamefont {Chernyshev}},\
  }\href@noop {} {\bibfield  {journal} {\bibinfo  {journal} {Phys. Rev. X}\
  }\textbf {\bibinfo {volume} {9}},\ \bibinfo {pages} {021017} (\bibinfo {year}
  {2019})}\BibitemShut {NoStop}%
\bibitem [{\citenamefont {Goodenough}(1963)}]{goodenough1963magnetism}%
  \BibitemOpen
  \bibfield  {author} {\bibinfo {author} {\bibfnamefont {J.~B.}\ \bibnamefont
  {Goodenough}},\ }\href@noop {} {\emph {\bibinfo {title} {Magnetism and the
  chemical bond}}},\ Vol.~\bibinfo {volume} {1}\ (\bibinfo  {publisher}
  {Interscience publishers},\ \bibinfo {year} {1963})\BibitemShut {NoStop}%
\bibitem [{\citenamefont {Ma}\ \emph {et~al.}(2016)\citenamefont {Ma},
  \citenamefont {Kamiya}, \citenamefont {Hong}, \citenamefont {Cao},
  \citenamefont {Ehlers}, \citenamefont {Tian}, \citenamefont {Batista},
  \citenamefont {Dun}, \citenamefont {Zhou},\ and\ \citenamefont
  {Matsuda}}]{ma2016static}%
  \BibitemOpen
  \bibfield  {author} {\bibinfo {author} {\bibfnamefont {J.}~\bibnamefont
  {Ma}}, \bibinfo {author} {\bibfnamefont {Y.}~\bibnamefont {Kamiya}}, \bibinfo
  {author} {\bibfnamefont {T.}~\bibnamefont {Hong}}, \bibinfo {author}
  {\bibfnamefont {H.}~\bibnamefont {Cao}}, \bibinfo {author} {\bibfnamefont
  {G.}~\bibnamefont {Ehlers}}, \bibinfo {author} {\bibfnamefont
  {W.}~\bibnamefont {Tian}}, \bibinfo {author} {\bibfnamefont {C.}~\bibnamefont
  {Batista}}, \bibinfo {author} {\bibfnamefont {Z.}~\bibnamefont {Dun}},
  \bibinfo {author} {\bibfnamefont {H.}~\bibnamefont {Zhou}}, \ and\ \bibinfo
  {author} {\bibfnamefont {M.}~\bibnamefont {Matsuda}},\ }\href@noop {}
  {\bibfield  {journal} {\bibinfo  {journal} {Physical review letters}\
  }\textbf {\bibinfo {volume} {116}},\ \bibinfo {pages} {087201} (\bibinfo
  {year} {2016})}\BibitemShut {NoStop}%
\bibitem [{\citenamefont {Ito}\ \emph {et~al.}(2017)\citenamefont {Ito},
  \citenamefont {Kurita}, \citenamefont {Tanaka}, \citenamefont
  {Ohira-Kawamura}, \citenamefont {Nakajima}, \citenamefont {Itoh},
  \citenamefont {Kuwahara},\ and\ \citenamefont {Kakurai}}]{ito2017structure}%
  \BibitemOpen
  \bibfield  {author} {\bibinfo {author} {\bibfnamefont {S.}~\bibnamefont
  {Ito}}, \bibinfo {author} {\bibfnamefont {N.}~\bibnamefont {Kurita}},
  \bibinfo {author} {\bibfnamefont {H.}~\bibnamefont {Tanaka}}, \bibinfo
  {author} {\bibfnamefont {S.}~\bibnamefont {Ohira-Kawamura}}, \bibinfo
  {author} {\bibfnamefont {K.}~\bibnamefont {Nakajima}}, \bibinfo {author}
  {\bibfnamefont {S.}~\bibnamefont {Itoh}}, \bibinfo {author} {\bibfnamefont
  {K.}~\bibnamefont {Kuwahara}}, \ and\ \bibinfo {author} {\bibfnamefont
  {K.}~\bibnamefont {Kakurai}},\ }\href@noop {} {\bibfield  {journal} {\bibinfo
   {journal} {Nature communications}\ }\textbf {\bibinfo {volume} {8}},\
  \bibinfo {pages} {1} (\bibinfo {year} {2017})}\BibitemShut {NoStop}%
\bibitem [{\citenamefont {Ghioldi}\ \emph {et~al.}(2018)\citenamefont
  {Ghioldi}, \citenamefont {Gonzalez}, \citenamefont {Zhang}, \citenamefont
  {Kamiya}, \citenamefont {Manuel}, \citenamefont {Trumper},\ and\
  \citenamefont {Batista}}]{ghioldi2018dynamical}%
  \BibitemOpen
  \bibfield  {author} {\bibinfo {author} {\bibfnamefont {E.~A.}\ \bibnamefont
  {Ghioldi}}, \bibinfo {author} {\bibfnamefont {M.~G.}\ \bibnamefont
  {Gonzalez}}, \bibinfo {author} {\bibfnamefont {S.-S.}\ \bibnamefont {Zhang}},
  \bibinfo {author} {\bibfnamefont {Y.}~\bibnamefont {Kamiya}}, \bibinfo
  {author} {\bibfnamefont {L.~O.}\ \bibnamefont {Manuel}}, \bibinfo {author}
  {\bibfnamefont {A.~E.}\ \bibnamefont {Trumper}}, \ and\ \bibinfo {author}
  {\bibfnamefont {C.~D.}\ \bibnamefont {Batista}},\ }\href@noop {} {\bibfield
  {journal} {\bibinfo  {journal} {Physical Review B}\ }\textbf {\bibinfo
  {volume} {98}},\ \bibinfo {pages} {184403} (\bibinfo {year}
  {2018})}\BibitemShut {NoStop}%
\bibitem [{\citenamefont {Kamiya}\ \emph {et~al.}(2018)\citenamefont {Kamiya},
  \citenamefont {Ge}, \citenamefont {Hong}, \citenamefont {Qiu}, \citenamefont
  {Quintero-Castro}, \citenamefont {Lu}, \citenamefont {Cao}, \citenamefont
  {Matsuda}, \citenamefont {Choi}, \citenamefont {Batista}, ,\ and\
  \citenamefont {Mourigal}}]{kamiya2018nature}%
  \BibitemOpen
  \bibfield  {author} {\bibinfo {author} {\bibfnamefont {Y.}~\bibnamefont
  {Kamiya}}, \bibinfo {author} {\bibfnamefont {L.}~\bibnamefont {Ge}}, \bibinfo
  {author} {\bibfnamefont {T.}~\bibnamefont {Hong}}, \bibinfo {author}
  {\bibfnamefont {Y.}~\bibnamefont {Qiu}}, \bibinfo {author} {\bibfnamefont
  {D.}~\bibnamefont {Quintero-Castro}}, \bibinfo {author} {\bibfnamefont
  {Z.}~\bibnamefont {Lu}}, \bibinfo {author} {\bibfnamefont {H.}~\bibnamefont
  {Cao}}, \bibinfo {author} {\bibfnamefont {M.}~\bibnamefont {Matsuda}},
  \bibinfo {author} {\bibfnamefont {E.}~\bibnamefont {Choi}}, \bibinfo {author}
  {\bibfnamefont {C.}~\bibnamefont {Batista}}, , \ and\ \bibinfo {author}
  {\bibfnamefont {M.}~\bibnamefont {Mourigal}},\ }\href@noop {} {\bibfield
  {journal} {\bibinfo  {journal} {Nature communications}\ }\textbf {\bibinfo
  {volume} {9}},\ \bibinfo {pages} {1} (\bibinfo {year} {2018})}\BibitemShut
  {NoStop}%
\bibitem [{\citenamefont {Sanders}\ \emph {et~al.}(2021)\citenamefont
  {Sanders}, \citenamefont {Mole}, \citenamefont {Liu}, \citenamefont {Brown},
  \citenamefont {Yu}, \citenamefont {Ling},\ and\ \citenamefont
  {Rachel}}]{sanders2021dominant}%
  \BibitemOpen
  \bibfield  {author} {\bibinfo {author} {\bibfnamefont {A.~L.}\ \bibnamefont
  {Sanders}}, \bibinfo {author} {\bibfnamefont {R.~A.}\ \bibnamefont {Mole}},
  \bibinfo {author} {\bibfnamefont {J.}~\bibnamefont {Liu}}, \bibinfo {author}
  {\bibfnamefont {A.~J.}\ \bibnamefont {Brown}}, \bibinfo {author}
  {\bibfnamefont {D.}~\bibnamefont {Yu}}, \bibinfo {author} {\bibfnamefont
  {C.~D.}\ \bibnamefont {Ling}}, \ and\ \bibinfo {author} {\bibfnamefont
  {S.}~\bibnamefont {Rachel}},\ }\href@noop {} {\bibfield  {journal} {\bibinfo
  {journal} {arXiv preprint arXiv:2112.12254}\ } (\bibinfo {year}
  {2021})}\BibitemShut {NoStop}%
\bibitem [{\citenamefont {Samarakoon}\ \emph {et~al.}(2021)\citenamefont
  {Samarakoon}, \citenamefont {Chen}, \citenamefont {Zhou},\ and\ \citenamefont
  {Garlea}}]{samarakoon2021static}%
  \BibitemOpen
  \bibfield  {author} {\bibinfo {author} {\bibfnamefont {A.~M.}\ \bibnamefont
  {Samarakoon}}, \bibinfo {author} {\bibfnamefont {Q.}~\bibnamefont {Chen}},
  \bibinfo {author} {\bibfnamefont {H.}~\bibnamefont {Zhou}}, \ and\ \bibinfo
  {author} {\bibfnamefont {V.~O.}\ \bibnamefont {Garlea}},\ }\href@noop {}
  {\bibfield  {journal} {\bibinfo  {journal} {Physical Review B}\ }\textbf
  {\bibinfo {volume} {104}},\ \bibinfo {pages} {184415} (\bibinfo {year}
  {2021})}\BibitemShut {NoStop}%
\bibitem [{\citenamefont {Chaloupka}\ and\ \citenamefont
  {Khaliullin}(2016)}]{chaloupka2016magnetic}%
  \BibitemOpen
  \bibfield  {author} {\bibinfo {author} {\bibfnamefont {J.}~\bibnamefont
  {Chaloupka}}\ and\ \bibinfo {author} {\bibfnamefont {G.}~\bibnamefont
  {Khaliullin}},\ }\href@noop {} {\bibfield  {journal} {\bibinfo  {journal}
  {Phys. Rev. B}\ }\textbf {\bibinfo {volume} {94}},\ \bibinfo {pages} {064435}
  (\bibinfo {year} {2016})}\BibitemShut {NoStop}%
\bibitem [{\citenamefont {Huyan}\ \emph {et~al.}(2022)\citenamefont {Huyan},
  \citenamefont {Schmidt}, \citenamefont {Gati}, \citenamefont {Zhong},
  \citenamefont {Cava}, \citenamefont {Canfield},\ and\ \citenamefont
  {Bud'ko}}]{huyan2022hydrostatic}%
  \BibitemOpen
  \bibfield  {author} {\bibinfo {author} {\bibfnamefont {S.}~\bibnamefont
  {Huyan}}, \bibinfo {author} {\bibfnamefont {J.}~\bibnamefont {Schmidt}},
  \bibinfo {author} {\bibfnamefont {E.}~\bibnamefont {Gati}}, \bibinfo {author}
  {\bibfnamefont {R.}~\bibnamefont {Zhong}}, \bibinfo {author} {\bibfnamefont
  {R.~J.}\ \bibnamefont {Cava}}, \bibinfo {author} {\bibfnamefont {P.~C.}\
  \bibnamefont {Canfield}}, \ and\ \bibinfo {author} {\bibfnamefont {S.~L.}\
  \bibnamefont {Bud'ko}},\ }\href@noop {} {\bibfield  {journal} {\bibinfo
  {journal} {arXiv preprint arXiv:2201.12233}\ } (\bibinfo {year}
  {2022})}\BibitemShut {NoStop}%
\bibitem [{\citenamefont {Georges}\ \emph {et~al.}(2013)\citenamefont
  {Georges}, \citenamefont {Medici},\ and\ \citenamefont
  {Mravlje}}]{georges2013strong}%
  \BibitemOpen
  \bibfield  {author} {\bibinfo {author} {\bibfnamefont {A.}~\bibnamefont
  {Georges}}, \bibinfo {author} {\bibfnamefont {L.~d.}\ \bibnamefont {Medici}},
  \ and\ \bibinfo {author} {\bibfnamefont {J.}~\bibnamefont {Mravlje}},\
  }\href@noop {} {\bibfield  {journal} {\bibinfo  {journal} {Annu. Rev.
  Condens. Matter Phys.}\ }\textbf {\bibinfo {volume} {4}},\ \bibinfo {pages}
  {137} (\bibinfo {year} {2013})}\BibitemShut {NoStop}%
\end{thebibliography}%

\appendix
\section{Low-Energy Multiplet Composition}
\label{sec:appendixA}

The pure $L,S$ multiplets $|m_L, m_S\rangle$ can be conveniently expressed in terms of the single-particle levels with precise orbital momentum:
\begin{align}
|e_{a,\sigma}\rangle =& \  |d_{z^2,\sigma}\rangle
\\
|e_{b,\sigma}\rangle =& \  |d_{x^2-y^2,\sigma}\rangle
\\
|t_{+,\sigma}\rangle = & \ -\frac{1}{\sqrt{2}}\left(|d_{yz,\sigma}\rangle + i |d_{xz,\sigma}\rangle \right)
\\
|t_{0,\sigma}\rangle = & \ |d_{xy,\sigma}\rangle
\\
|t_{-,\sigma}\rangle = & \ \frac{1}{\sqrt{2}}\left(|d_{yz,\sigma}\rangle - i |d_{xz,\sigma}\rangle \right)
\end{align}
This leads to:
\begin{align}
\left|-1,\frac{3}{2}\right\rangle =  & \ \left|e_{a,\uparrow}e_{b,\uparrow}t_{+,\uparrow}t_{0,\uparrow}t_{0,\downarrow}t_{-,\uparrow}t_{-,\downarrow}\right\rangle 
\\
\left|0,\frac{1}{2}\right\rangle = & \ \frac{1}{\sqrt{3}} \left|e_{a,\uparrow}e_{b,\uparrow}t_{+,\uparrow}t_{+,\downarrow}t_{0,\downarrow}t_{-,\uparrow}t_{-,\downarrow}\right\rangle 
 \\
& \ 
+\frac{1}{\sqrt{3}}   \left|e_{a,\uparrow}e_{b,\downarrow}t_{+,\uparrow}t_{+,\downarrow}t_{0,\uparrow}t_{-,\uparrow}t_{-,\downarrow}\right\rangle 
\nonumber \\
& \ 
+\frac{1}{\sqrt{3}}  \left|e_{a,\downarrow}e_{b,\uparrow}t_{+,\uparrow}t_{+,\downarrow}t_{0,\uparrow}t_{-,\uparrow}t_{-,\downarrow}\right\rangle 
\nonumber \\
\left|1,-\frac{1}{2}\right\rangle = & \  \frac{1}{\sqrt{3}} \left|e_{a,\uparrow}e_{b,\downarrow}t_{+,\uparrow}t_{+,\downarrow}t_{0,\uparrow}t_{0,\downarrow}t_{-,\downarrow}\right\rangle 
\\
& \ 
+\frac{1}{\sqrt{3}} \left|e_{a,\downarrow}e_{b,\uparrow}t_{+,\uparrow}t_{+,\downarrow}t_{0,\uparrow}t_{0,\downarrow}t_{-,\downarrow}\right\rangle 
\nonumber \\
& \ 
+\frac{1}{\sqrt{3}} \left|e_{a,\downarrow}e_{b,\downarrow}t_{+,\uparrow}t_{+,\downarrow}t_{0,\uparrow}t_{0,\downarrow}t_{-,\uparrow}\right\rangle 
\nonumber 
\end{align}
The time-reversed partners can be similarly obtained. 

\end{document}